\documentstyle[pre,aps]{revtex} 
\begin{document}
\draft
\title{Critical phenomena of nonequilibrium dynamical systems 
with two absorbing states} 
\author{WonMuk Hwang
\thanks{Address after September 1, 1998: Department of Physics,
Boston University, Boston, MA 02215.}}
\address{Department of Physics, Korea Military Academy, Seoul 139-799, Korea}
\author{Sungchul Kwon, Heungwon Park, and Hyunggyu Park
\thanks{Author to whom correspondence should be addressed. 
Electronic address: hgpark@munhak.inha.ac.kr}}
\address{Department of Physics, Inha University,  Inchon 402-751, Korea}

\date{\today}
\maketitle            
\begin{abstract}
We study nonequilibrium dynamical models with two absorbing states: 
interacting monomer-dimer models, probabilistic cellular automata models,
nonequilibrium kinetic Ising models. These models exhibit a continuous
phase transition from an active phase into an absorbing phase which belongs 
to the universality class of the models with the parity
conservation. However, when we break the symmetry between the
absorbing states by introducing a symmetry-breaking field, Monte Carlo simulations
show that the system goes back to the conventional 
directed percolation universality class. In terms of domain wall language, the
parity conservation is not affected by the presence of the symmetry-breaking field.
So the symmetry between the absorbing states rather than the conservation
laws plays an essential role
in determining the universality class. We also
perform Monte Carlo simulations for the various interface dynamics between different
absorbing states, which yield new universal dynamic exponents.
With the symmetry-breaking field, the interface moves, in average,
with a constant velocity in the direction of the unpreferred absorbing state and
the dynamic scaling exponents apparently assume trivial values.  However, we
find that the hyperscaling relation for the directed percolation universality class is
restored if one focuses on the dynamics of the interface on the side of the
preferred absorbing state only. 
\end{abstract}

\section{Introduction} 
Many nonequilibrium dynamical models show continuous phase transitions similar to
ordinary equilibrium  models. In  fact, nonequilibrium  models  can supply  much richer 
critical behavior because their evolving dynamics do not require the detailed balance. 
So the universality classes of nonequilibrium critical phenomena would be much more
diverse and  would  be governed   by various  symmetry properties   of the  evolution 
dynamics. 

An interesting example of nonequilibrium phase transitions is the absorbing phase
transition. In this case,  there exist some absorbing  states in the configurational  phase 
space. If the system gets into one of the absorbing states by the evolution dynamics, 
then the system is trapped inside of the absorbing states and no further dynamics occur to
escape out of the absorbing states.
By controlling an external parameter, one can observe a 
continuous phase transition from an active steady-state phase into an inactive absorbing
phase. Recently, various kinds of nonequilibrium models exhibiting such an absorbing
phase transition  have been  studied extensively 
\cite{marro}.  Most of  the models  investigated are 
found to belong to the directed percolation (DP) universality class 
\cite{GrassTorre,CardySugar,Janssen,Grass82,Grin,r,t,j1,ko,u,u1}.
A common feature  of these models  is that  the absorbing phase  consists of a  single 
absorbing state. 

Only a few models have been studied that are not in the DP universality class.
Those are the models A and B of probabilistic cellular automata (PCA)
\cite{Grassberger1,Grassberger2}, 
nonequilibrium kinetic Ising models with two different dynamics (NKI)
\cite{Menyhard1,Menyhard2,Menyhard3}, and interacting 
monomer-dimer models (IMD)
\cite{KimPark,ParkKimPark}. Numerical investigations show that critical behaviors of
these models are different from DP but form a non-DP universality class. These models 
share a common property that the absorbing phase consists of two equivalent absorbing 
states. By the analogy to the equilibrium Ising model which has two equivalent 
ground states, we call this non-DP universality
class as the {\em directed Ising} (DI) universality class. 

Recently, the branching  annihilating random walks  (BAW) with  offsprings have been 
studied intensively \cite{Bram,Sud,Taka,Taka2,Jen934,Jen931,Jen941,Redner,Kwon}
Even though the BAW model has a single absorbing state
(vacuum), its critical behavior depends on the parity of the number of offsprings. 
It has been shown numerically that the 
BAW models with an odd number of offsprings (BAWo) belong to the DP class, while
the BAW models with an even number of offsprings (BAWe) belong to the DI class 
\cite{Taka,Jen941}. 
Dynamics of the BAWe  models conserve the  number of walkers modulo  2, while the 
BAWo models evolve without any conservation. The common feature of the PCA, NKI,
IMD, and BAWe models is that the number of particles (walkers in BAWe and kinks or
domain walls in the other models) is conserved modulo 2. From this point of view, 
it was suggested that the parity conservation is responsible for the DI universality 
class. This is why the DI universality class is sometimes called as the PC 
(parity-conserving) universality class. 

However, we recently showed for the IMD model that an external
field which conserves the parity but breaks the symmetry between two absorbing states 
forces the system back to the conventional DP universality class \cite{ParkPark}. 
So we argued that the symmetry between absorbing states rather than the conservation
laws plays an essential role in determining the universality class. 
Our argument was supported by recent results 
for generalized monomer-monomer models studied by Bassler and Browne 
\cite{Bassler1,Bassler2,Bassler3},
and for some stochastic models by Hinrichsen \cite{Hinrichsen}.

In this paper, we study the effect of a symmetry-breaking field in the IMD, PCA, and NKI
models via stationary as well as dynamic Monte Carlo simulations. Stationary simulations 
and defect dynamics for all three models clearly show that the DI universality class
crosses over to the DP class
under a weak parity-conserving symmetry-breaking field. In fact, the ratio of
the number of stationary runs which fall into the unpreferred absorbing state 
and the number of those into the preferred state 
vanishes exponentially in system size.
So the system with the symmetry-breaking field has in effect
a single absorbing state, which leads to the DP class.

We also introduce new types of interface dynamics which result in different
values of dynamic scaling exponents. These exponents are found to be universal.
Without a symmetry-breaking field, 
the hyperscaling relation for the DI universality class is intact 
for various interface dynamics.
However, with the symmetry-breaking field, the interface moves, in average,
with a constant velocity in the direction of the unpreferred absorbing state. 
The dynamic scaling exponents apparently assume trivial values and violate the hyperscaling
relation.  However, we find that the hyperscaling relation 
for the DP universality class is restored if one focuses on the dynamics of 
the interface on the side of the preferred absorbing state only. 

In the next section, we  report our numerical results for the IMD model 
in various dynamic simulations.
In Sec.~III, the effect of the symmetry-breaking field in the IMD model
is dicussed in details via stationary and dynamic simulations.
In Sec.~IV, the numerical results for the PCA and NKI models are presented. 
Finally we conclude in Sec.~V with summary and discussion.


\section{Dynamic critical behavior of the IMD model}

The interacting monomer-dimer (IMD) model is a generalization of the simple
monomer-dimer model on a catalytic surface, in which particles of the same
species have nearest-neighbor repulsive interactions \cite{KimPark}. Here 
we  consider the one-dimensional IMD model with infinitely
strong repulsions between the same species. 
A monomer ($A$) cannot adsorb at a nearest-neighbor site of an already-occupied
monomer (restricted vacancy) but adsorbs at a free vacant site with no adjacent 
monomer-occupied sites. Similarly, a dimer ($B_2$) cannot adsorb at a 
pair of restricted vacancies ($B$ in nearest-neighbor sites) but adsorbs at a 
pair of free vacancies. There are no nearest-neighbor restrictions in 
adsorbing particles of different species. Only the adsorption-limited reactions 
are considered. Adsorbed dimers dissociate and a nearest neighbor 
of the adsorbed $A$ and $B$ particles reacts,
forms the $AB$ product, and desorbs the catalytic surface immediately.
Whenever there is an $A$ adsorption attempt at a vacant site between 
an adsorbed $A$ and an adsorbed $B$, we allow the $A$ to adsorb and react
immediately with the neighboring $B$, thus forming the $AB$ product and
desorbing the surface. 
If this process is not allowed, the IMD model possesses
infinitely many absorbing states and is found to be always absorbing
\cite{HwangPark}. 

The system has no fully 
saturated phases of monomers or dimers, but
instead two equivalent half-filled absorbing states. These states are
comprised of only the monomers at the odd- or even-numbered lattice sites,
i.e.~$(A0A0\cdots)$ and $(0A0A\cdots)$ where ``$A$'' represents a 
monomer-occupied site and ``$0$'' a vacant site. These two states are
probabilistically equivalent, unless we introduce a symmetry-breaking field
discriminating the dynamics at the odd- and even-numbered sites. In this 
section,
we consider the IMD model without a symmetry-breaking field. 
This model can be parametrized by the monomer adsorption-attempt probability
$p$. The dimer adsorption-attempt probability is then given by $1-p$.

In our previous stationary and dynamic Monte Carlo simulations
\cite{KimPark,ParkKimPark}, it was found that 
the system undergoes a continuous phase transition from a reactive phase
into an absorbing phase, which belongs to the DI universality class. 
The kink representation of the IMD model is complicated due to
its multi-component nature. 
Three types of kinks can be defined between lattice sites
occupied by a dimer and a dimer, by a dimer and a vacancy, and
by a vacancy and a vacancy. No conservation law 
is associated with each type of kinks,
but the total number of kinks is conserved modulo 2. So, in a broad sense,
one can say that the IMD model evolves by the parity-conserving dynamics
like the BAWe model. In this section, we discuss the dynamic critical 
behavior of the IMD model via Monte Carlo simulations. Some of the results
reported previously \cite{ParkKimPark} are much improved using new 
efficient algorithms
and some new results for various interface dynamics are presented.

\subsection{Defect dynamics}

We start with a 
lattice occupied by monomers at alternating sites except at the central
vacant site, i.e. $(\cdots A0A0{\underline 0}0A0A0\cdots)$ where ${\underline 0}$
represents a defect at the central site. Then the system evolves along the dynamic
rule of the model. 
After one adsorption attempt on the average per lattice 
site (one Monter Carlo time step), the time is incremented by one unit.
A number of independent runs, typically $5 \times 10^5$, are made up to $3\times 10^4$
time steps for various values of $p$ near the critical probability $p_c$.
Most runs, however,  stop earlier because the system enters into 
one of the absorbing states.
We measure the survival probability
$P(t)$ (the probability that the system is still active  at time t), the number
of dimers $N(t)$ averaged over all runs,  and the mean square
distance of spreading of the active region  $R^2 (t)$ averaged over surviving
runs. At criticality, the values of these quantities  scale algebraically in the
long time limit \cite{GrassTorre},  
\begin{eqnarray}
P(t)&\sim t^{-\delta}\label{Pt}\nonumber\\
N(t)&\sim t^\eta\label{Nt}\\
R^2 (t)  &\sim t^z\label{R2t}\nonumber  
\end{eqnarray}
and double-logarithmic plots of these values against time show straight 
lines at critiality. Off criticality, these plots show curvatures. 
More precise estimates for the dynamic scaling exponents can be obtained by examining 
the local slopes of the curves. The effective exponent $\delta(t)$ is defined 
as
\begin{equation}
-\delta(t) = {{\log[P(t)/P(t/b)]}\over{\log(b)}} \label{eff}
\end{equation}
and similary for $\eta(t)$ and $z(t)$.  
In this paper, we plot 
the effective exponents against $10/t$ with $b=10$.
Off criticality, these plots 
show positive or negative curvatures. The scaling exponents can be extracted
by taking the asymptotic values of the effective exponents at criticality.

Our estimates for the critical probability 
and the dynamic scaling exponets are $p_c =0.5322(3)$, $\delta=0.290(5)$, 
$\eta=0.00(1)$, and  $z=1.135(5)$ (see Fig.~1). Note that the estimate for $z$ is
much improved compared with our previous result; $z=1.34(20)$ \cite{ParkKimPark}.
These values are in excellent agreement with those of the 
DI universality class such as the BAWe model \cite{Jen941}.

\subsection{Interface dynamics}

For the interface dynamics, we start with a pair of vacancies
placed at the central sites of a lattice and with monomers occupied at
alternating sites, i.e.~$(\cdots A0A{\underline {00}}A0A\cdots)$ where
the interface between two different absorbing states is placed 
in the middle of two central vacancies ${\underline {00}}$. 
In this case, the system never enters an absorbing state, so that 
the survival probability is always equal to 1 and the exponent $\delta=0$.
Even though the values of $\delta$ and $\eta$ vary with the types of dynamics,
their sum $\delta +\eta$ which is responsible for the growth of the number 
of kinks (or dimers) in surviving runs is known to be universal 
\cite{Mendes,ParkKimPark}. This guarantees
that the generalized hyperscaling relation is always satisfied;
$\beta/\nu_\parallel +\delta +\eta = dz/2$ where $d$ is the spatial dimension
and $\beta$ and $\nu_\parallel$ are steady-state exponents explained in the
next section. As $\delta=0$ in 
this type of interface dynamics, it does not supply any new information about
dynamic critical behavior of the system. 

In this section, we introduce three different types of interface dynamics 
which may give nontrivial scaling of the survivial probability $P(t)$. For 
convenience, the ordinary interface dynamics as above is called as 
{\em type-A interface dynamics}. Our previous study for the type-A
interface dynamics found that $\eta=0.285(20)$ and $z=1.14(2)$ as expected 
\cite{ParkKimPark}.

In the {\em type-B interface dynamics}, we stop the evolution if the
interface collapses to its initial configuration, i.e.~two vacant sites 
between the absorbing states. Then this run is treated as a dead one. 
This dynamics is originally introduced by
Bassler and Browne for a three species monomer-monomer model \cite{Bassler1}. 
At criticality, we measure $P(t)$, $N(t)$, and $R^2(t)$. 
$P(t)$ now represents not a true survival probability but
a probability of avoiding a collapse. In Fig.~2, we plot the effective exponents
against $10/t$. Our estimates for the dynamic scaling exponents are
$\delta=0.73(1)$, $\eta=-0.41(1)$ and $z=1.16(2)$. These values satisfy the
generalized hyperscaling relation of the DI universality class 
and are in excellent accord with those 
reported by Bassler and Browne \cite{Bassler1}. 
To see how much these exponents are robust,
we change the criteria for stopping the evolution  
from two active sites to four and six active sites. 
We find that there is no essential change in
the values of the exponents. Therefore the nontrivial value of $\delta$ 
in the type-B dynamics is believed to be universal. 

In the {\em type-C interface dynamics}, we focus only on the
profile between the central site and the leftmost site (the left front) 
of the active region. 
We stop the evolution when the left front of the active region comes back to the
center (initial position) and treat this run as a dead one (see Fig.~3). 
So $P(t)$ represents a probability of avoiding a collapse of the
active region in the left side with respect to the initial location of the interface. 
We measure $N(t)$ as the number of dimers only in the left side of the center and
$R^2 (t)$ as the mean square spreading of the active region in the left side of the
center. The type-C interface dynamics is useful when one needs to distinguish 
the behavior of the left and right fronts of the interface. So it is 
especially important
to consider this type of dynamics for the interface between unequivalent absorbing
states (see Sec.~III). The effective exponents against $10/t$ at 
criticality are plotted in Fig.~4. Our estimates are $\delta=0.395(5)$, $\eta=-0.10(1)$
and $z=1.150(5)$, which also satisfy the generalized hyperscaling relation of the DI
universality class.
The value of $\delta$ is different from those for the defect and other interface
dynamics. It would be useful to check that this exponent is also universal 
for other models in 
the DI universality class. In Sec.~IV, we study the type-C dynamics for the 
PCA model and find this exponent is universal. 

Finally, we introduce the {\em type-D interface dynamics.} This dynamics is similar
to the type-C dynamics in measuring physical quantities. But we do not 
stop the evolution when the left front of the active region hits the center. 
However, while its left front wanders in the right side of the center,
we treat this run as a dead one temporarily, and set $N(t)$ and $R^2 (t)$ to be zero.
When it comes back to the left side of the center, we treat this run as a surviving
run again and measure $N(t)$ and $R^2 (t)$ as usual in the type-C dynamics.
$P(t)$ represents a probability that the active region covers the left side of
the center and is expected to converge to a nonzero constant less than 1. 
The effective exponents against $10/t$ at 
criticality are plotted in Fig.~5. Our estimates are $\delta=0.010(5)$, 
$\eta=0.29(1)$
and $z=1.15(1)$. As expected, $\delta$ is nearly zero and $P(t)$ converges to 
0.69(1) (see Fig.~6).
So the type-D interface dynamics yields the same exponents as in the type-A
(ordinary) dynamics. The type-A dynamics does not yield the correct scaling
exponents for the interface between unequivalent absorbing states, so 
in this case the type-D dynamics can be employed instead (see Sec.~III-C).

\section{The IMD model with a symmetry-breaking field}

We introduce a symmetry-breaking field which makes the system to
prefer one absorbing state to the other \cite{ParkPark}. 
This can be done by differentiating the 
monomer adsorption-attempt probability $p$ at an odd-numbered vacant site and at 
an even-numbered one. If a monomer attempts to adsorb on an even-numbered
free vacant site, the adsorption attempt is rejected with probability
$h$ $(0 \le h \le 1)$. The case $h=0$ corresponds to the ordinary IMD model
discussed in the previous section. For finite $h$, the monomers tend to adsorb 
more on an odd-numbered site than an even-numbered one. So the absorbing
state with odd-numbered sites occupied by monomers is probabilistically 
preferable to the other absorbing state. However, the kink dynamics of this model
still conserves the parity in terms of the total number of kinks. 
In this section, we show that 
the symmetry-breaking field forces the system back to the conventional
DP universality class via stationary and dynamic simulations. Therefore one can
conclude that not the parity conservation but the symmetry between the absorbing 
states is essential in determining the universality class of the absorbing
phase transitions. 

\subsection{Stationary simulations}

We run stationary Monte Carlo simulations starting with an empty
lattice with size $L$. Then the system evolves along the dynamic rule
with the symmetry-breaking field using periodic boundary conditions. 
We set the value of the
symmetry-breaking field $h=0.5$ for convenience.
After a sufficiently long time, the system reaches a quasisteady state first and
stays for a reasonably long time before finally entering into an absorbing state. 
We measure the concentration of dimers in the quasisteady state and average over
many independent runs which have not yet entered into an absorbing state. 
The number of independent runs varies from
$5\times 10^4$ for the system size $L=32$ to $3\times 10^3$ 
for $L=512$. 

Elementary scaling theory combined with the finite-size scaling theory 
\cite{GrassTorre,Ankrust}
predicts that the average concentration of dimers ${\bar \rho}$
at criticality in the steady state scales with system size $L$ as
\begin{equation}
\bar \rho (L)\sim L^{-\beta/\nu_\bot}, \label{fss2}
\end{equation}
where $\beta$ is the order parameter exponent and 
$\nu_\bot$ is the correlation length exponent in the spatial direction. 
In the reactive phase ($p<p_c$), the concentration $\bar\rho$ remains
finite in the limit $L\rightarrow\infty$, but it should vanish exponentially 
with system size in
the absorbing phase ($p>p_c$). 

At $p_c$, we expect the ratio of the concentrations of
dimers for two successive system sizes
$\bar\rho(L/2)/\bar\rho(L)=2^{\beta/\nu_\bot}$, ignoring corrections to scaling.
This ratio converges to 1 for $p<p_c$ and approaches 2 for $p>p_c$
in the limit $L\rightarrow\infty$. We plot the logarithm of this ratio divided
by $\log 2$ as a function of $p$ for $L=64, 128, 256$ and 512 in Fig.~7.
The crossing points between lines for two successive sizes 
converge to the point at $p_c \simeq 0.413(1)$ and
$\beta/\nu_\bot \simeq 0.23(2)$. The critical probability 
can be more accurately estimated 
from the defect dynamics simulations; $p_c=0.4138(3)$ 
(see the next subsection). We run stationary simulations at 
this value of $p$ and find the better estimate for        
$\beta/\nu_\bot = 0.243(8)$ (see Fig.~8). This value is consistent with
the standard DP value of 0.2524(5) \cite{Jen941}. 

By analyzing the decay characteristics of dimer concentrations
at criticality, we can 
extract information about the relaxation time exponent $\nu_\parallel$.
From the elementary scaling theory, one can expect the short time behavior 
of the dimer concentration at criticality as
\begin{equation}
\bar\rho (t) \sim t^{-\beta/\nu_\|}.\label{fss5}
\end{equation}
The characteristic (relaxation) time $\tau$ for a finite system is defined as the 
elapsing time for a finite system to enter into the quasisteady state.
Then one can find $\tau$ scales at criticality as
\begin{equation}
\tau \sim L^{\nu_\| /\nu_\bot}.\label{fss6}
\end{equation}

In Fig.~9, we plot 
$\bar\rho$ at $p_c$ versus time for various system sizes. 
Investigating the slopes in this double-logarithmic plot, we estimate 
$\beta/\nu_\parallel = 0.165(10)$. 
The double-logarithmic plot (Fig.~10) for the characteristic time $\tau$ versus
the system size $L$ shows a straight line from which
we obtain $\nu_\|/\nu_\bot =1.45(10)$ which is consistent with the above results
and agrees reasonably well with the standard DP value of 1.580(2).

Similar results are obtained for the system with a weaker symmetry-breaking field. 
Therefore we conclude that the symmetry-breaking field in the system with
two absorbing states is relevant in determining the universality class 
and makes the system to behave like having a single absorbing state.  
In fact, the number of stationary runs falling into the
unpreferred absorbing state compared with the number of those into the
preferred state vanishes exponentially in system size (see Sec.~IV-A). 
The preferred absorbing
state behaves as a unique absorbing state of the system. 

\subsection{Defect dynamics}   

When the symmetry between the absorbing states is broken, the defect
dynamics are sensitive to the initial configurations. 
One can start with a lattice with a defect, either in the preferred absorbing
state or in the unpreferred absorbing state. 
The latter case does not show critical spreading of the active region.
The domain of the preferred absorbing state grows at the center of
the active region while the region of the unpreferred absorbing state 
recedes with a constant speed. 
This dynamics is much like two interface dynamics
between the preferred and unpreferred absorbing states
where two interfaces move in the opposite direction. We will discuss 
the interface dynamics in the next subsection. 

We choose the initial configuration with a defect at the center 
in the preferred absorbing state. We set $h=0.5$ for convenience. 
Our estimates for the critical probability and the
dynamic scaling exponents are
$p_c=0.4138(3)$, $\delta=0.163(5)$, $\eta=0.315(5)$, and $z=1.265(5)$ (see Fig.~11).
These values are also improved compared with our previous results \cite{ParkPark}
and are in excellent accord with the standard DP values;
$\delta=0.1596(4)$, $\eta=0.3137(10)$, and $z=1.2660(14)$.

\subsection{Interface dynamics}

As the two absorbing states are not probabilistically equivalent, 
the dynamics of the interface 
between the preferred and unpreferred absorbing states 
shows a completely different behavior from the symmetric case. 
In Fig.~12, evloutions of the interface below, at, and above criticality
are shown. 
For $p<p_c$, the active region grows linearly in time in both directions.
Of course, the left interface front near the preferred absorbing state
(we call it {\em P-interface front})
moves slower than the right interface front 
({\em U-interface front}) near the unpreferred absorbing state
due to the symmetry-breaking field. 
At the critical point, the P-interface front behaves like a critical interface
(zero velocity with nontrivial temporal scaling)
as in the critical defect dynamics, but the U-interface front
still moves with a finite velocity. 
As we increase $p$ further into the absorbing phase, the 
active region cannot grow and the preferred absorbing state 
dominates. The P-interface front moves with the same velocity and
the same direction as the U-interface front. The width of the active region
is finite, in contrast to its diffusive behavior in the absorbing phase
for the symmetric case. Here, we consider the critical case only.

In the symmetric case, four different interface dynamics are introduced.
In the ordinary (type-A) interface dynamics, both of the exponents
$\eta$ and $z/2$ converge to a trivial value of unity, due to the
ballistic nature of the U-interface front even at criticality. 
The type-B interface dynamics also involve the dynamics of the U-interface 
front, so it does not yield nontrival values of the exponents. The type-C and 
type-D interface dynamics focus only on the profile between the
initial position of the interface (the center) and the P-interface front which 
behaves in a critical fashion. 

For the type-C dynamics, we run $2\times 10^4$ independent samples 
up to $3\times 10^4$
time steps at $p_c=0.4138$ with $h=0.5$. Our estimates are
$\delta=0.37(1)$, $\eta=0.11(1)$, and $z=1.27(1)$ (Fig.~13).
As expected, the value of $z$ agrees well with the DP value and 
these exponents satisfiy the generalized hyperscaling relation 
for the DP universality class, i.e.~$\delta +\eta = 0.48(2)$ 
is in  excellent accord with the standard DP value of 0.473(1).

Similarly, we find  
$\delta=0.02(2)$, $\eta=0.45(1)$, and $z=1.23(2)$ for the type-D dynamics 
(Fig.~14). Again these values agree well with the DP values. 
The concept of concentrating only on the relevant interface front 
may be applied to other types of models in which many types of interfaces 
coexist and some of them are not equivalent.

\section{Other models}

\subsection{Probabilistic celluar automata }

Grassberger, Krause, and von der Twer \cite{Grassberger1,Grassberger2}
studied two models of probabilistic cellular automata (PCA), 
namely models A and B about ten years ago. 
The A model evolves with rule number 94 in the notation of Wolfram 
\cite{Wolfram}
except $110$ and $011$ configurations, where the central spin 1
flips to 0 with probability $p$ and remains unflipped with
probability $1-p$. The B model evolves with rule number 50
except $110$ and $011$ configurations, where the central spin 1
flips to 0 with probability $1-p$ and remains unflipped with
probability $p$. 
These are the first models investigated which are not in the DP
universality class.
Both models have two equivalent absorbing states,
i.e.~$(1010\cdots)$ and $(0101\cdots)$, and 
exhibit an absorbing phase transition which belongs to the
DI universality class. But these models behave differently in the absorbing
phase. Once the system enters into one of the two absorbing states,
it remains in that state forever in model A but oscillates from one state
to the other in model B. In spite of the discriminating behavior, these
models belong to the same DI universality class. In the kink representation, 
the total number of kinks are conserved modulo 2 in the dynamics. 

First, consider the A model. We can introduce the symmetry-breaking field 
which makes the system to prefer $(1010\cdots)$ to $(0101\cdots)$. 
The system must go through a $111$ configuration right before entering 
into an absorbing state, i.e.~$(\cdots
010\underline{111}010\cdots) ~\rightarrow ~ (\cdots
010\underline{101}010\cdots)$. If the flipping probability of the
central spin in the $111$ configuration depends on the even- or oddness
of the position of the central spin, the symmetry between two absorbing states 
can be broken. With probability $h$, we reject the flipping attempt of the central 
spin in the $111$ configuration when it is at an even-numbered site. 
Then the absorbing state with 1's at the odd-numbered sites is probabilistically
preferable to the other one. Again the parity in the total
number of kinks is conserved even with the symmetry-breaking field $h$. 

We run the defect and type-C interface dynamics with $h=0.1$. Our estimates 
for the defect dynamics are 
$p_c=0.2435(4)$, $\delta=0.1625(25)$, $\eta=0.310(5)$,
and $z=1.245(5)$ (Fig.~15). 
These values agree very well with the standard DP values.
For the type-C interface dynamics at criticality,
we find $\delta=0.372(2)$, $\eta=0.115(5)$, and $z=1.275(5)$ (Fig.~16), 
which agree with the results
for the IMD model with the symmetry-breaking field (see Sec.~III-C). 
So the value of $\delta$ in the type-C dynamics seems to be universal. 
In order to check whether the value of $\delta$ in the type-C dynamics 
is universal in the symmetric case, we run the type-C interface dynamics
when $h=0$ and find $\delta=0.395(5)$, $\eta=-0.08(2)$, and $z=1.17(2)$
(Fig.~17), which is consistent
with those found in the IMD model without the symmetry-breaking field
(see Sec.~II-B). 

We also run
stationary simulations at criticality with size $L=32$ up to 512 with $h=0.1$.
Our estimates for the steady-state exponents are
$\beta/\nu_\bot = 0.245(5)$, $\beta/\nu_\parallel = 0.155(5)$, and 
$\nu_\parallel/\nu_\bot = 1.63(5)$ (Fig.~18) which also agree reasonably
well with the DP values. We measure the 
number of stationary runs falling into 
the unpreferred and preferred absorbing states 
respectively, i.e. $N_u$ and $N_p$ at criticality in the long time limit.
The ratio $R=N_u / N_p$ versus system size $L$ is plotted
in Fig.~19. This ratio vanishes exponentially in system size;
$R \sim \exp(-L/L_0)$ with $L_0 \simeq 13$.
It means that the chance of entering into the unpreferred absorbing 
state is neglible so the system behaves like having a single 
absorbing state of the preferred one.

For the B model, the situation is quite different. The system 
oscillates between the two absorbing states; $(1010\cdots)
\leftrightarrow (0101\cdots)$. 
In this model, the system must go through a $000$ configuration 
right before entering into an absorbing state, i.e.~~$(\cdots
101\underline{000}101\cdots) ~\rightarrow ~ (\cdots
010\underline{101}010\cdots)$. We can introduce 
a rejection probability discriminating the even- and odd-numbered sites
similar to that in the A model, but it cannot make the two absorbing states 
probabilistically unequivalent due to the oscillatory nature in the dynamics.
In fact, it is meaningless to distinguish the two absorbing states without
any dynamic barrier. In the A model, there exists a dynamic barrier 
between the two absorbing states, which makes the system to take 
an infinitely long time to hop from near one absorbing state to near
the other absorbing state. This dynamic barrier is similar to
the free energy barrier between two ground states in the equilibrium Ising model. 
Without this barrier, there is no way to distinguish the two absorbing states.
Therefore it seems impossible to find the crossover 
from the DI to the DP universality class in the B model.

\subsection{Nonequilibrium kinetic Ising model  }

Nonequilibrium kinetic Ising model (NKI) recently introduced by Menyh\'ard 
\cite{Menyhard1,Menyhard2,Menyhard3} evolves with the competing effect of spin flips  
at zero temperature $(T=0)$
and nearest-neighbor spin exchanges at $T=\infty$. 
The spin-flip dynamics occurs with probability $p$ and spin-exchange dynamics with 
$1-p$. 
The competition between the two different dynamics at different temperatures
drives the system into an nonequilibrium steady state and there is a continuous
phase transition as the competition parameter $p$ varies. 

In the $T=0$ spin-flip dynamics, the system evolves trying to lower
the energy. 
A spin is allowed to flip only if the flip lowers the energy of the
system or leaves it unchanged. 
We use the parameter $r$ to distinguish the cases when
the energy is lowered or unchanged. We flip a spin with probability
$r$ in the former case and flip a spin freely in the latter case.
Here, we set $r=0.5$.
In the $T=\infty$ spin-exchange dynamics, nearest-neighbor spins 
are freely exchanged regardless of the energy change. Any up-down pair
of spins can flip to the down-up pair of spins if they are in the nearest neighbor.

The absorbing phase consists of two completely 
ferromagnetically ordered states which are equivalent. 
One can not flip a spin in these absorbing states because it increases the energy. 
These absorbing states are the same as the two degenerate ground states of the
equilibrium ferromagnetic Ising model. 
The absorbing transition of the NKI model belongs to the DI universality class.
In terms of ordinary domain-wall language, 
a domain wall between different spins is interpreted as a walker in the BAW model.
Then the NKI model can be mapped exactly to the BAW model with two
offsprings with a control parameter for the two walker annihilation process
\cite{Redner,Parkun}.

We introduce a symmetry-breaking field which prefers up spins over down spins.
So this field plays like an external magnetic field in the equilibrium Ising model.
For convenience, we define the symmetry-breaking field $h$ as a probability of
not allowing a up spin to flip. Therefore the absorbing state with all spins up
becomes the preferred absorbing state. Again the parity in the total number of
domain walls is still conserved even with the symmetry-breaking field.

We run the defect dynamics and stationary simulations 
with $h=0.1$. Our estimates are
$p_c = 0.190(5)$, $\delta=0.17(1)$, $\eta=0.32(1)$, and $z=1.25(5)$ (Fig.~20),
which agree well with the standard DP value. 
Stationary simulations at criticality yield 
$\beta/\nu_\bot = 0.24(1)$, $\beta/\nu_\parallel = 0.155(5)$, and 
$\nu_\parallel/\nu_\bot = 1.50(5)$ (Fig.21) which also agree reasonably
well with the DP values.


\section{Summary and discussion}
                                           
All models studied in this paper (IMD, PCA, NKI)
preserve the parity of the total number of kinks but cross over from the 
directed Ising (DI) to
directed percolation (DP) universality class 
when the parity-conserving symmetry-breaking field is introduced. 
As we argued in our preliminary paper \cite{ParkPark}, 
the essential factor which determines the universality class of a 
nonequilibrium absorbing phase transition is not the conservation laws in dynamics
but the symmetry between absorbing states. 

We take a careful look at various kinds of kinks in these models. 
First, consider the NKI model which is the simplest one in the
domain wall (or kink) representation. In the NKI model,
a kink is assigned between two neighboring spins in the opposite direction.
Only one type of kinks exists in the NKI model and there is a
two-to-one mapping between spin configurations and kink configurations.
The two absorbing states correspond to the vacuum configuration in the
kink representation. 
The evolution dynamics conserves the total number of kinks modulo 2.
By identifying a kink as a walker in the BAW model, 
the NKI model can be exactly mapped 
to the BAW model with two offsprings 
with a control parameter for the two walker annihilation process \cite{Redner,Parkun}.
Numerical results \cite{Taka,Jen941}
and recent field theoretical works for the BAW models 
\cite{Cardy1,Cardy2}
suggest that the parity conservation is responsible for the DI universality
class. The symmetry-breaking field (magnetic field) in the NKI model
cannot be represented by a local kink operator in the field theory language,
similar to the magnetic field in the
equilibrium Ising model in the Bloch wall representation.
Therefore the recent field theoretical results by Cardy and T\"auber \cite{Cardy1,Cardy2} 
do not apply when the symmetry-breaking field is introduced.
In order to consider the symmetry-breaking field, one should include
the long-range string operator, i.e.~the global product of the number operators 
of kinks in the quantum Hamiltonian, which becomes a highly nontrivial problem. 
Our numerical results suggest that this long-range string operator is 
relevant and makes the system to leave a DI fixed point and 
flow into a DP fixed point by the renormalization group transformations.

The PCA models are similar to the NKI models even though 
there seems no trivial mapping into the BAW models. These models
contain two types of kinks which are assigned between two neighboring
1's and 0's respectively. There is no kink between 1 and 0, so the two
absorbing states correspond to the vacuum in the kink representation.
Due to the parallel updating procedure, it is not easy to examine
whether these two types of kinks can be represented by one kink
operator in the field theory. But the parity of the total number of
kinks is conserved during the evolution.
The symmetry-breaking field
which discriminates the even-numbered and odd-numered sites should
be also represented by a long-range string operator like that in the NKI model.
However, the B model does not have any dynamic barrier between two
absorbing states due to the oscillatory nature,
so is not affected by the symmetry-breaking field. 
Except that, we 
can draw the same conclusion for the PCA models as in the NKI model.

The IMD model has a more complex kink representation due to its
multi-component nature. Three different types of kinks are found between
$B$ and $B$, $B$ and 0, 0 and 0, where $B$ is a site
occupied by a dimer atom and 0 is a vacancy. The parity of the total number 
of kinks are conserved and the symmetry-breaking field is similar to
that in the PCA models. Our numerical results show that the IMD model 
exhibits the same critical behavior as in the NKI and PCA model. It may
imply that the differences between various kinks in the IMD and PCA models
are just irrelevant details which do not affect the universal behavior. 
It may be interesting to study these differences in the field theoretical
models.

Recently, a few other models have been introduced 
with equivalent absorbing states. 
Those are generalized monomer-monomer models
studied by Bassler and Browne 
\cite{Bassler1,Bassler2,Bassler3}, and generalized  Domany-Kinzel models and 
generalized contact processes studied by Hinrichsen \cite{Hinrichsen}. 
These models are multi-component models so there are many types of
kinks. Unlike the IMD model, there appears no explicit parity conservation law 
in the kink representation of these models, even though they all have 
two equivalent absorbing states. Numerical simulations showed that
these models belong to the DI universality class. 
By introducing a symmetry-breaking field, these models cross over from
the DI to DP universality class as in the models studied in this paper. 
These results strongly support our conclusion that the symmetry between 
absorbing states, not the conservation law, is the essential property to 
determine the universality class of the absorbing phase transitions. 
However, we do not exclude the possibility of the hidden conservation law
in the kink dynamics of these models. In their absorbing phase, 
numerical simulations show that large domains
of two different absorbing states are formed and active regions between two
different domains (domain walls with finite width) 
survive and diffuse until they annihilate pairwise \cite{Hinrichsen}. 
It implies that there may be an effective parity conservation law in domain walls,
even though there is no parity conservation in microscopic kinks.
With the symmetry-breaking field, large domains of the unpreferred absorbing state
completely disappear and domain walls (active regions) annihilate by themselves.

In this paper, we also introduce and investigate various interface dynamics.
Without the symmetry-breaking field, we find new universal exponents for $\delta$
in the type-B and type-C interface dynamics, but the hyperscaling relation
for the DI universality class
is always intact. With the symmetry-breaking field, the interface moves, in average,
with a constant velocity in the direction of the unpreferred absorbing state. 
By focusing only on the side of the preferred absorbing state, we
find new exponents for $\delta$ for the type-C and type-D dynamics
and the hyperscaling relation for the DP universality class is obtained.
These new exponents are also shown to be universal.  
These types of interface dynamics should be useful in studying some other models
with many but unequivalent absorbing states. 

It is interesting to compare the two dimensional equilibrium Ising model 
with the one dimensional NKI model. 
It is well known that two dimensional equilibrium models are related to
one dimensional kinetic models via transfer matrix formalism {\cite {dd}}. 
An extra space dimension is interpreted as the time dimension in one dimensional
kinetic models. One can write down the evolution operator of
the kinetic model corresponding to a given two dimensional equilibrium model.
This evolution operator is Hermitian for the equilibrium model.
General nonequilibrium kinetic models in one dimension 
can be obtained by modifying the
above Hermitian evolution operator in a non-Hermitian form, i.e.~breaking the
detailed balance. Then equilibrium models
in $d$ dimensions and nonequilibrium kinetic models in $d-1$ dimensions can be
directly compared. 

The NKI model is a special case
of general nonequilibrium kinetic Ising models. 
In the NKI model, the time reversal symmetry is 
broken completely and its dynamics favors one time direction over the other.
Therefore, by adding a directional sense in the time direction 
to the equilibrium Ising model,
one can see the crossover from the equilibrium Ising universality class to
the nonequilibrium DI universality class. 

Similar things happen for the models with a single absorbing state
which belong to the DP  universality class.
The percolation problem is equivalent to the $q\rightarrow 1$ limit of the
$q$-state Potts model \cite{g}. The DP problem is defined by 
adding a directional sense to the percolation problem.
Similarly, one can define nonequilibrium models with $q$ 
equivalent absorbing
states by adding a directional sense to the $q$-state Potts model. 
Both models have the permutation symmetry between $q$ ground
(or absorbing) states, so these nonequilibrium models may be called
as the $q$-state {\em directed} Potts model. 
In this sense, the NKI model may be called as the directed Ising model
and the NKI universality class as the directed Ising (DI) universality class.
It will be interesting to 
set up and investigate the directed Potts models 
with $q\ge 3$ \cite{Hinrichsen}. Of course, other types of
generalized models with the rotational or cyclic symmetry are also interesting.

There is a difference in connecting models with different values of $q$.
Crossover from the DI to the DP universality class is obtained by 
introducing a symmetry breaking field.
But the Ising phase transition disappears when the magnetic field is applied
to the two dimensional 
equilibrium Ising model. The percolation universality class appears only through
the random cluster formulation of the Potts model \cite{m}. 
Therefore the analogy between the $q$-state Potts model and the $q$-state 
directed Potts model is not complete. 
These similarities and differences should be further investigated in future.

\section*{acknowledgements}

This work is supported in part by  Korea Science
and Engineering Foundation through the SRC program of SNU-CTP and
by the academic research fund of Ministry of Education, Republic of Korea
(Grant No.~97-2409).


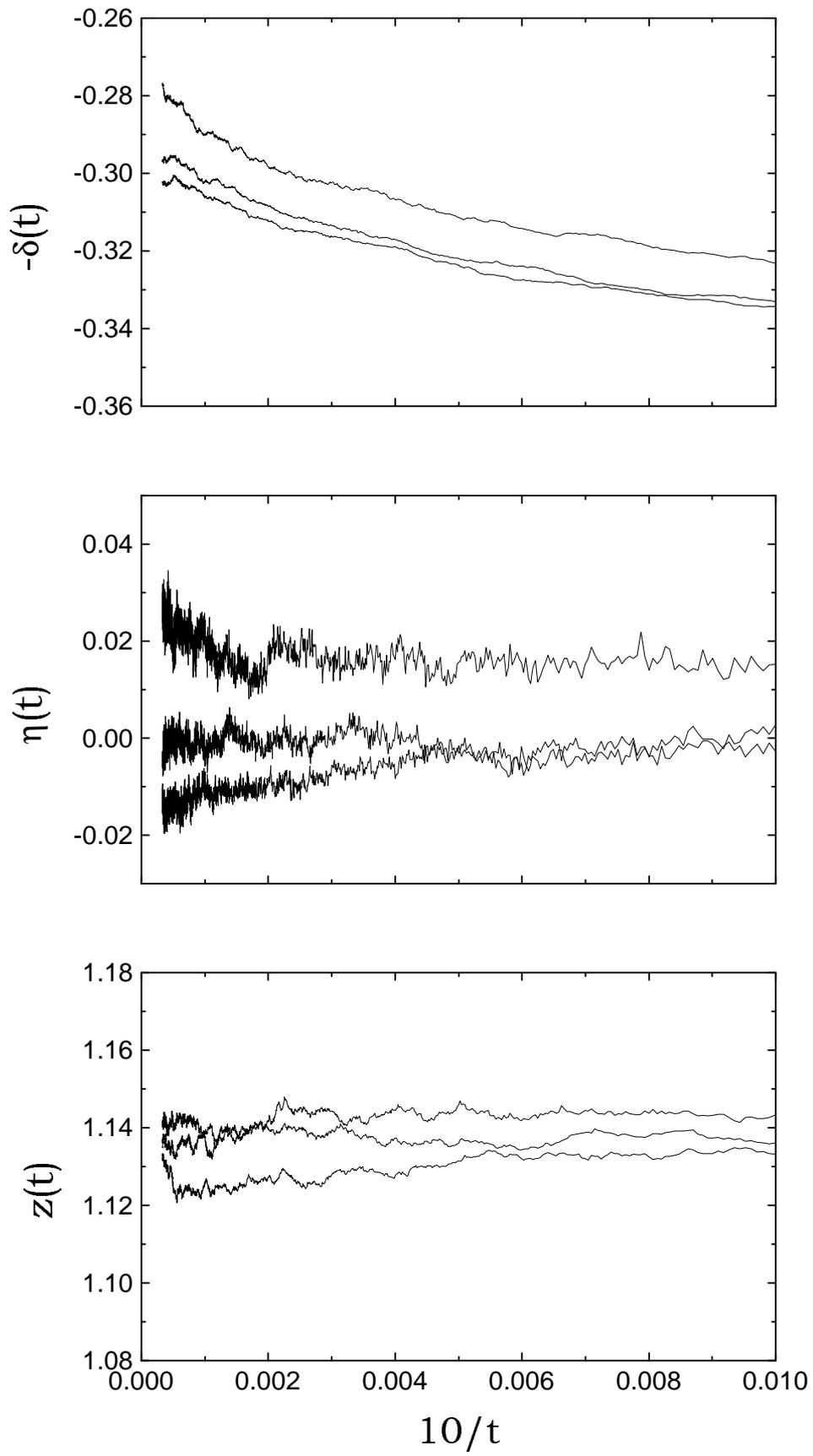
\begin{figure}
\caption{Plots of the effective exponents against $10/t$ for the 
defect dynamics of the symmetric IMD model.    
Three curves from top to bottom in each panel correspond to
$p=0.5315$, $0.5322$, and $0.5329$.}
\end{figure}

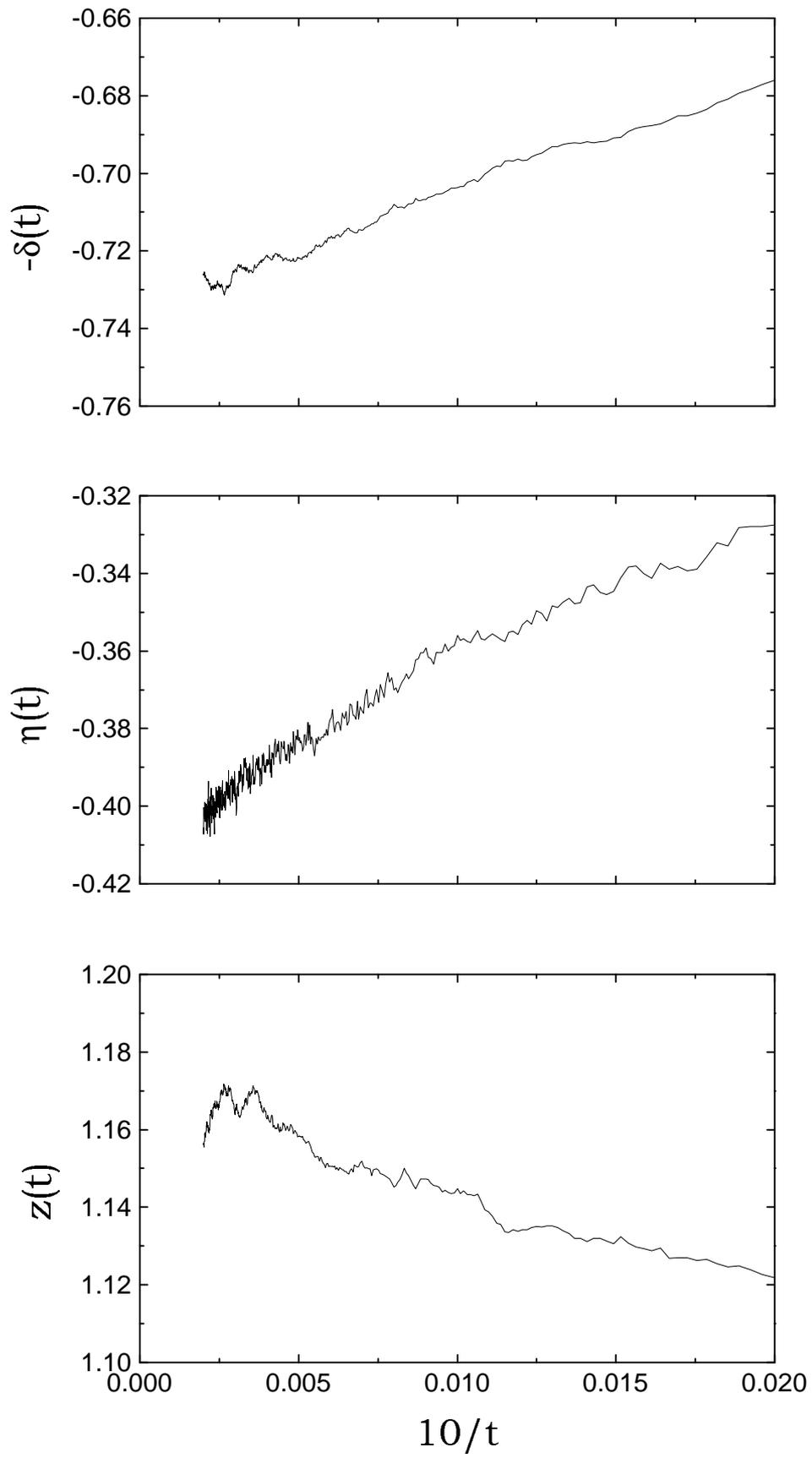
\begin{figure}
\caption{Plots of the effective exponents against $10/t$ for the 
type-B interface dynamics of the symmetric IMD model
at criticality $(p_c =0.5322)$.}
\end{figure}

\begin{figure}
\caption{The type-C interface dynamics in the IMD model. 
Grey, black, and white dots represent monomers, dimers, and vacancies, respectively. }
\end{figure}

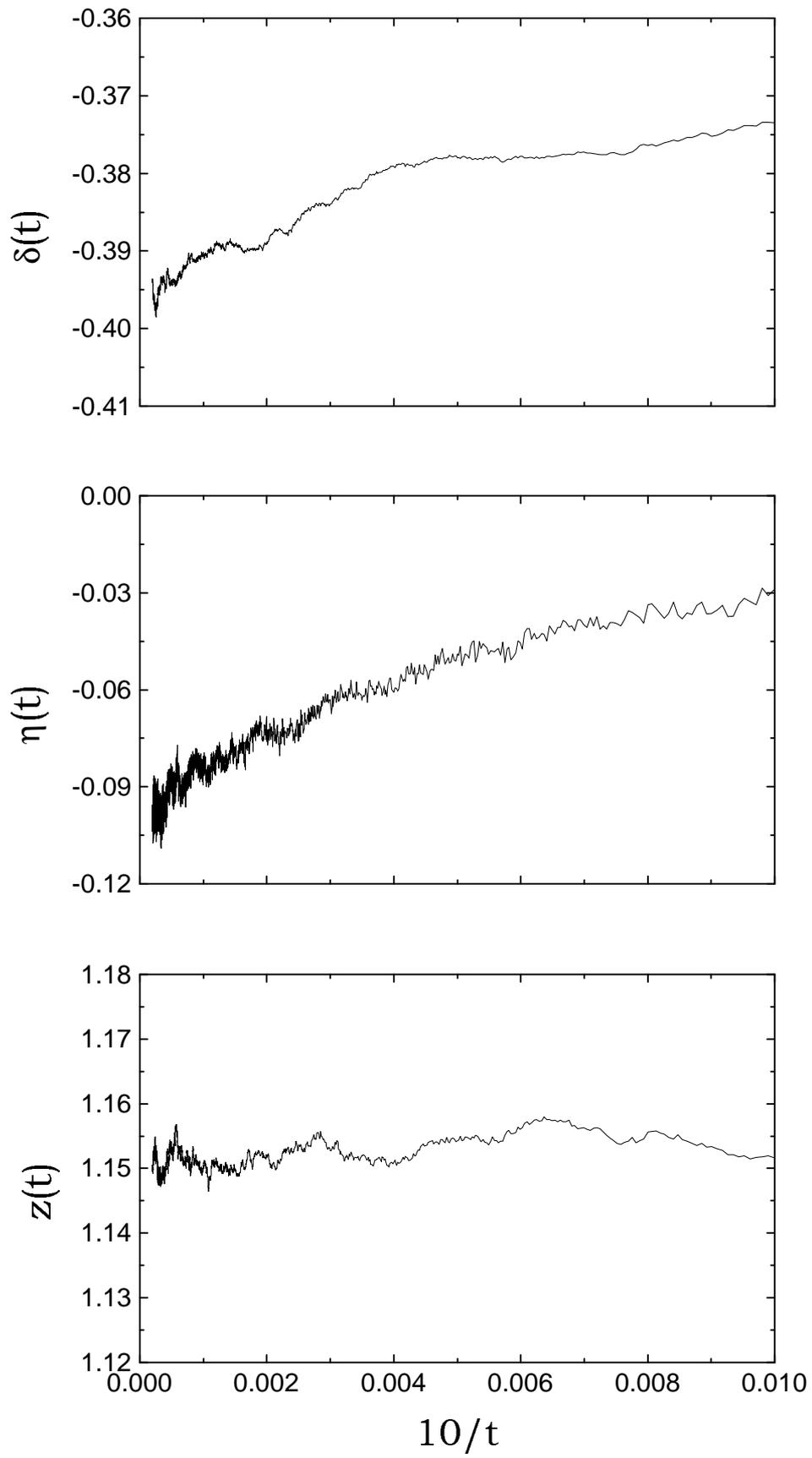
\begin{figure}
\caption{Plots of the effective exponents against $10/t$ for the 
type-C interface dynamics of the symmetric IMD model
at criticality.}
\end{figure}

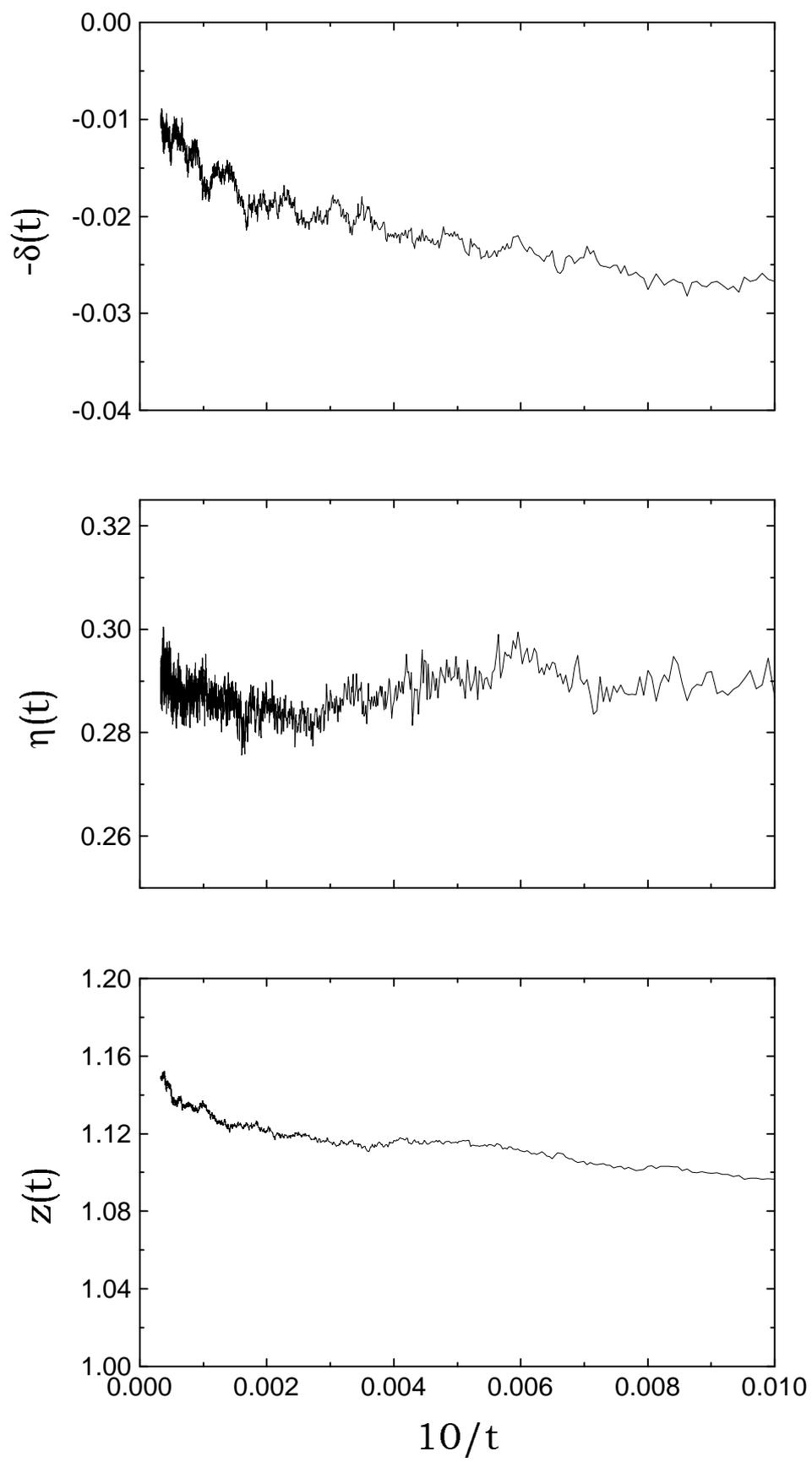
\begin{figure}
\caption{Plots of the effective exponents against $10/t$ for the 
type-D interface dynamics of the symmetric IMD model
at criticality.}
\end{figure}

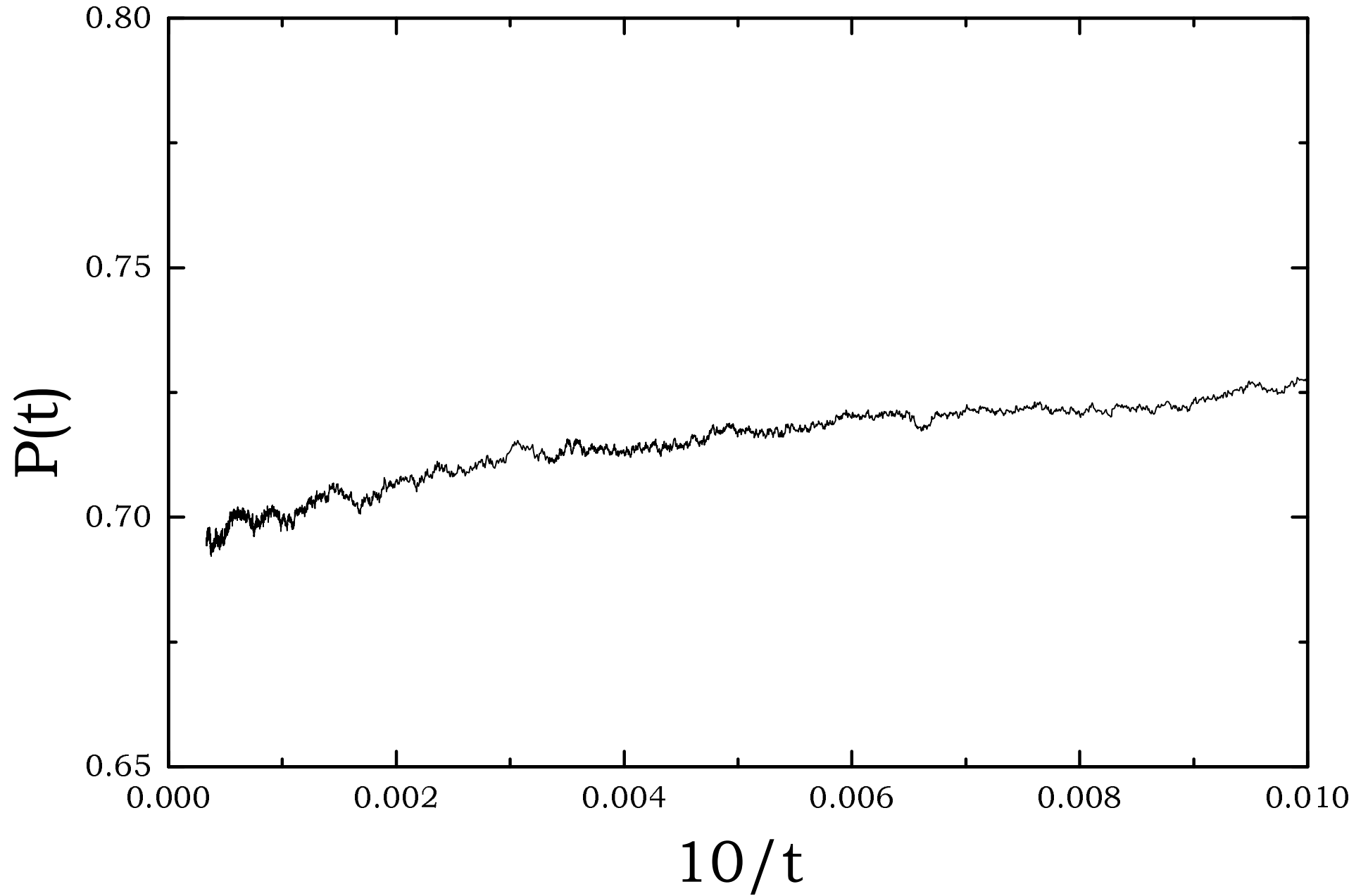
\begin{figure}
\caption{$P(t)$ versus $10/t$ for the type-D interface
dynamics of the symmetric IMD model.}
\end{figure}

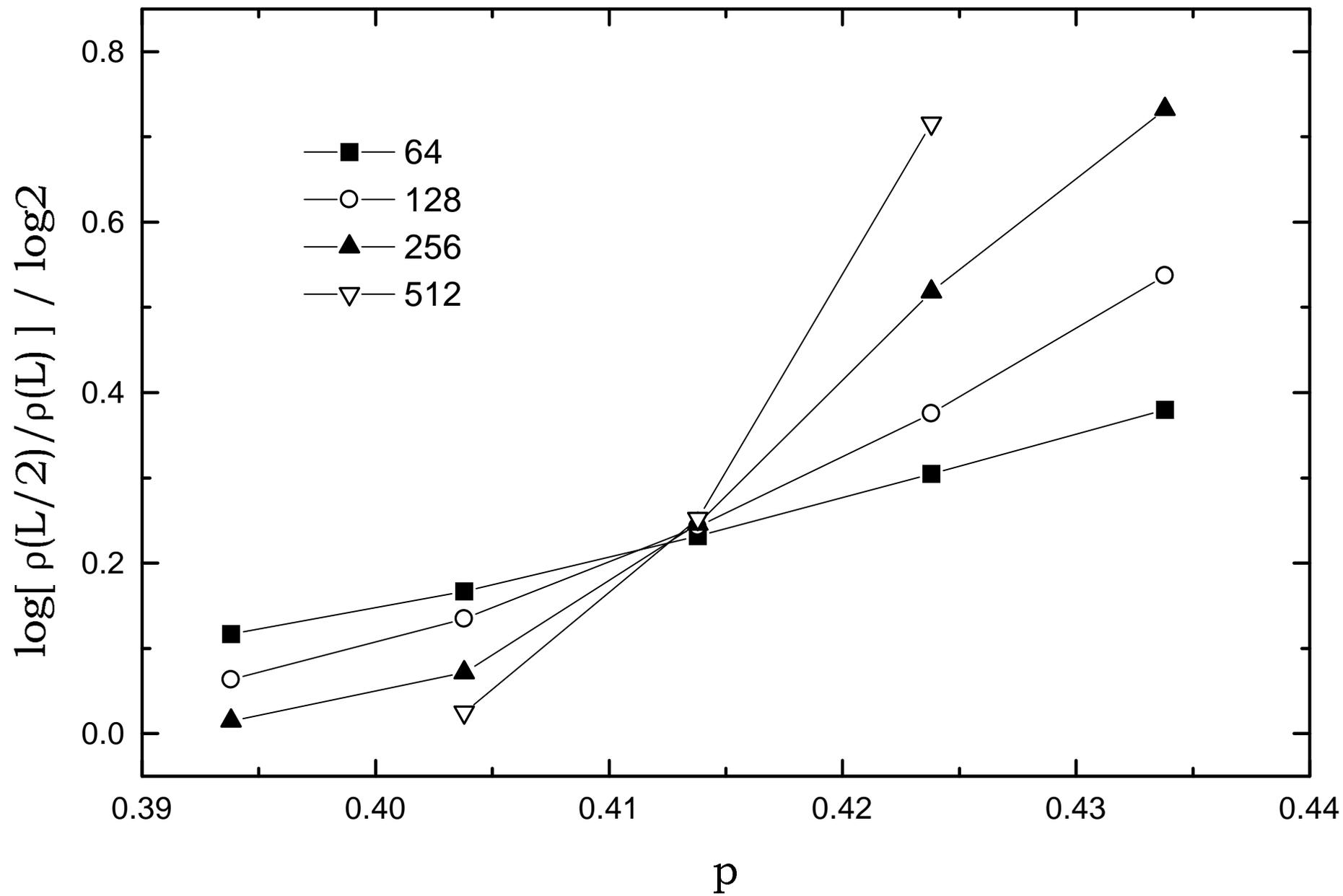
\begin{figure}
\caption{Plots of 
$\log\bigl[\bar\rho(L/2)/\bar\rho(L)\bigr]/\log(2)$ versus\ $p$ 
for the asymmetric IMD model with the symmetry-breaking field $h=0.5$
for various system sizes $L=64$, $128$, $256$, and $512$.}
\end{figure}

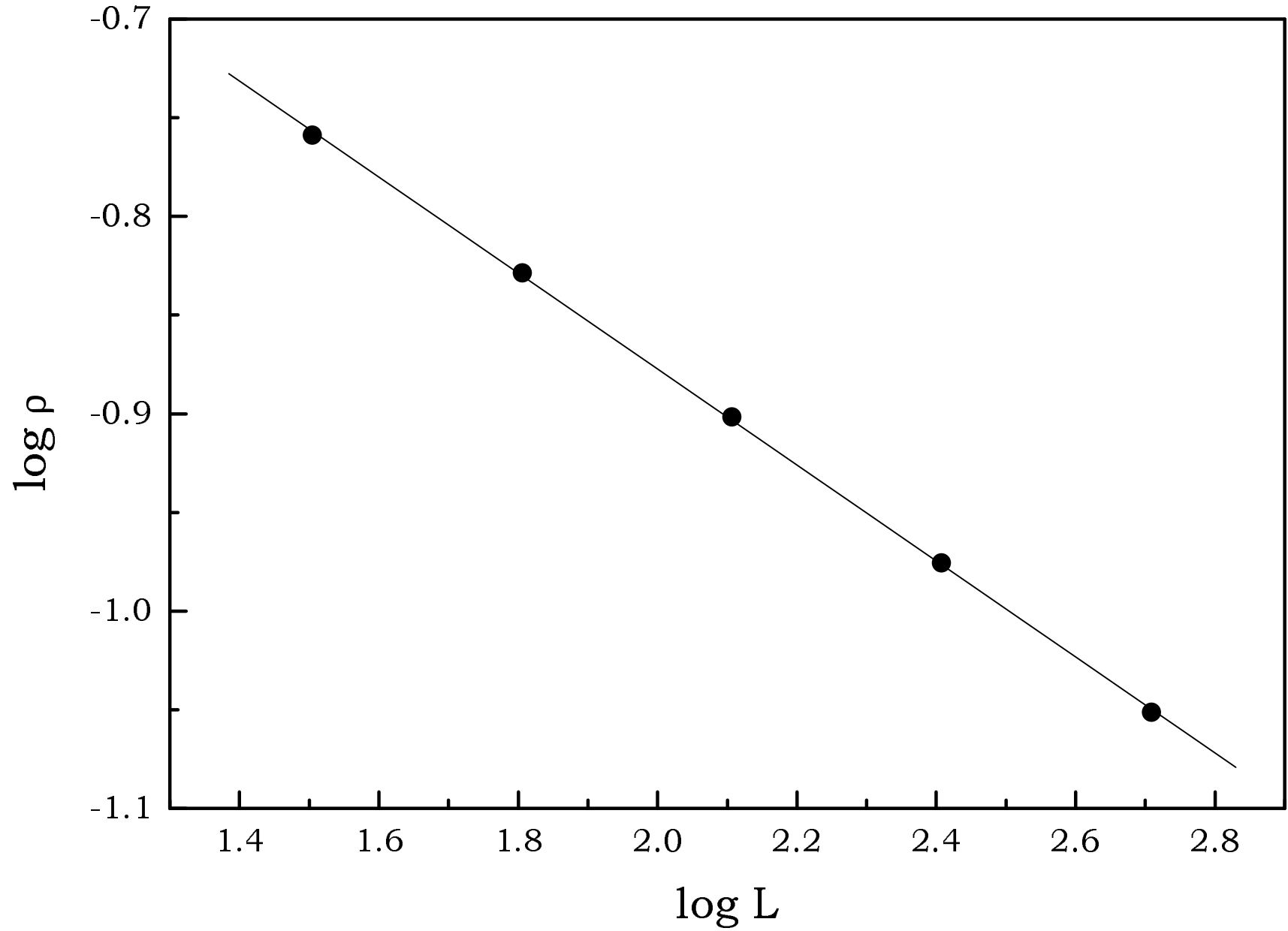
\begin{figure}
\caption{The average concentration of dimers $\bar\rho$ at criticality
in the quasisteady state against the system size $L$ in a double-logarithmic plots
for various system sizes $L=32 - 512$ for the asymmetric IMD model with $h=0.5$. 
The solid line is of slope $-0.243$.}                                  
\end{figure}

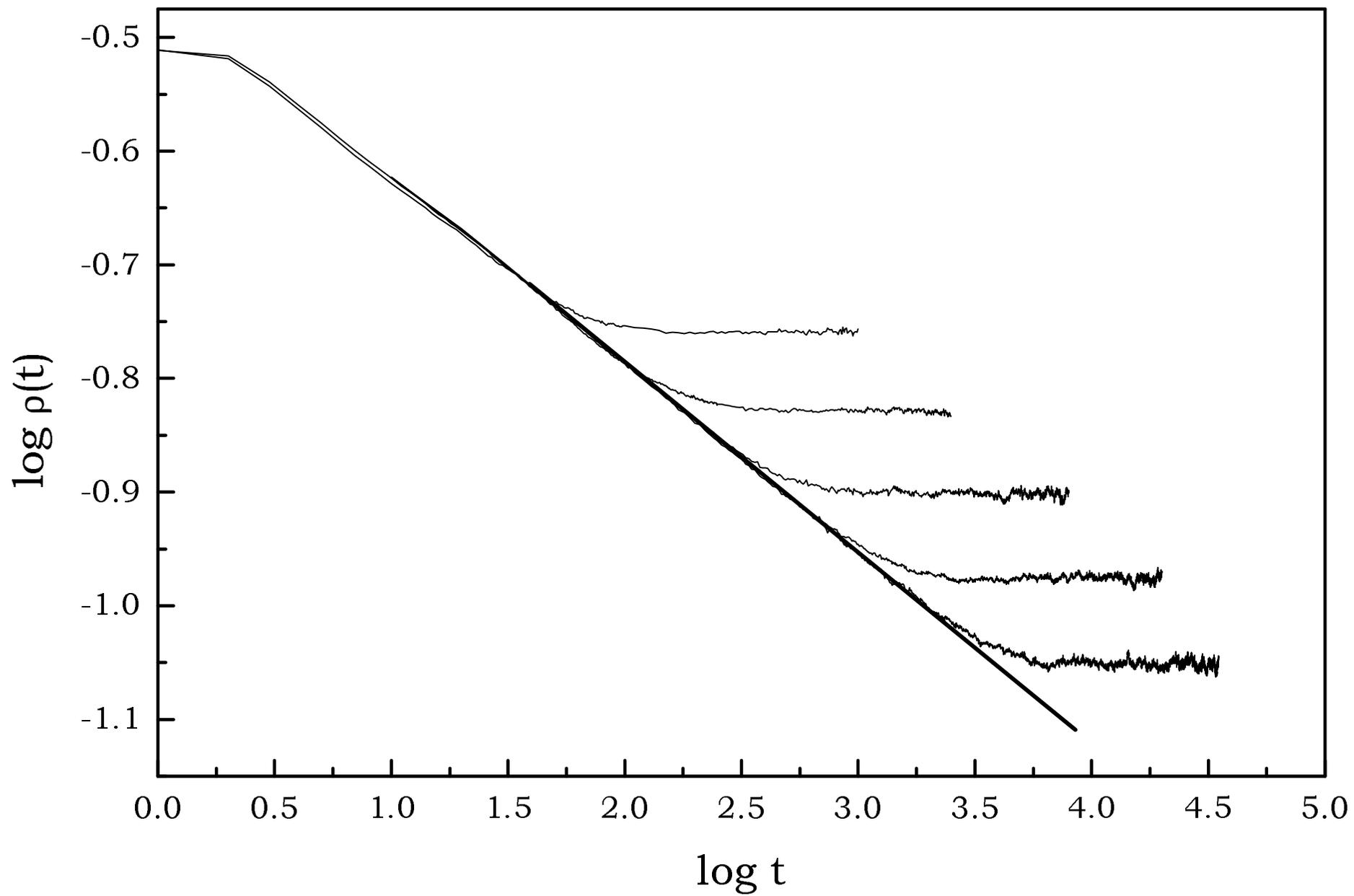
\begin{figure}
\caption{Time dependence of the average concentration of dimers at criticality
for various system sizes $L=32 - 512$ for the asymmetric IMD model with $h=0.5$. 
The solid line is of slope $-0.165$.}
\end{figure}

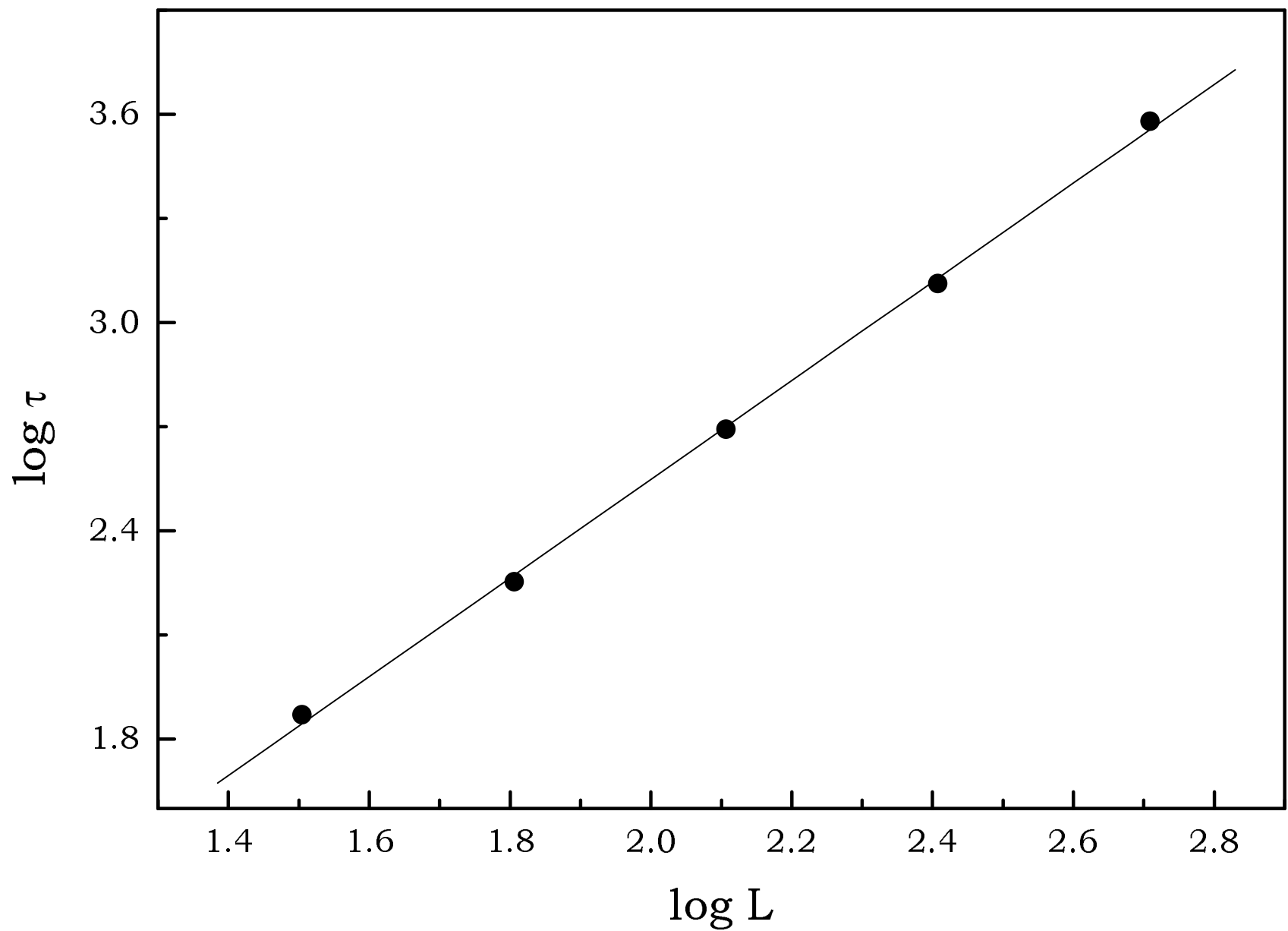
\begin{figure}
\caption{Size dependence of the characteristic time at criticality
for various system sizes $L=32 - 512$ for the asymmetric IMD model with $h=0.5$. 
The solid line is of slope $1.45$.} 
\end{figure}

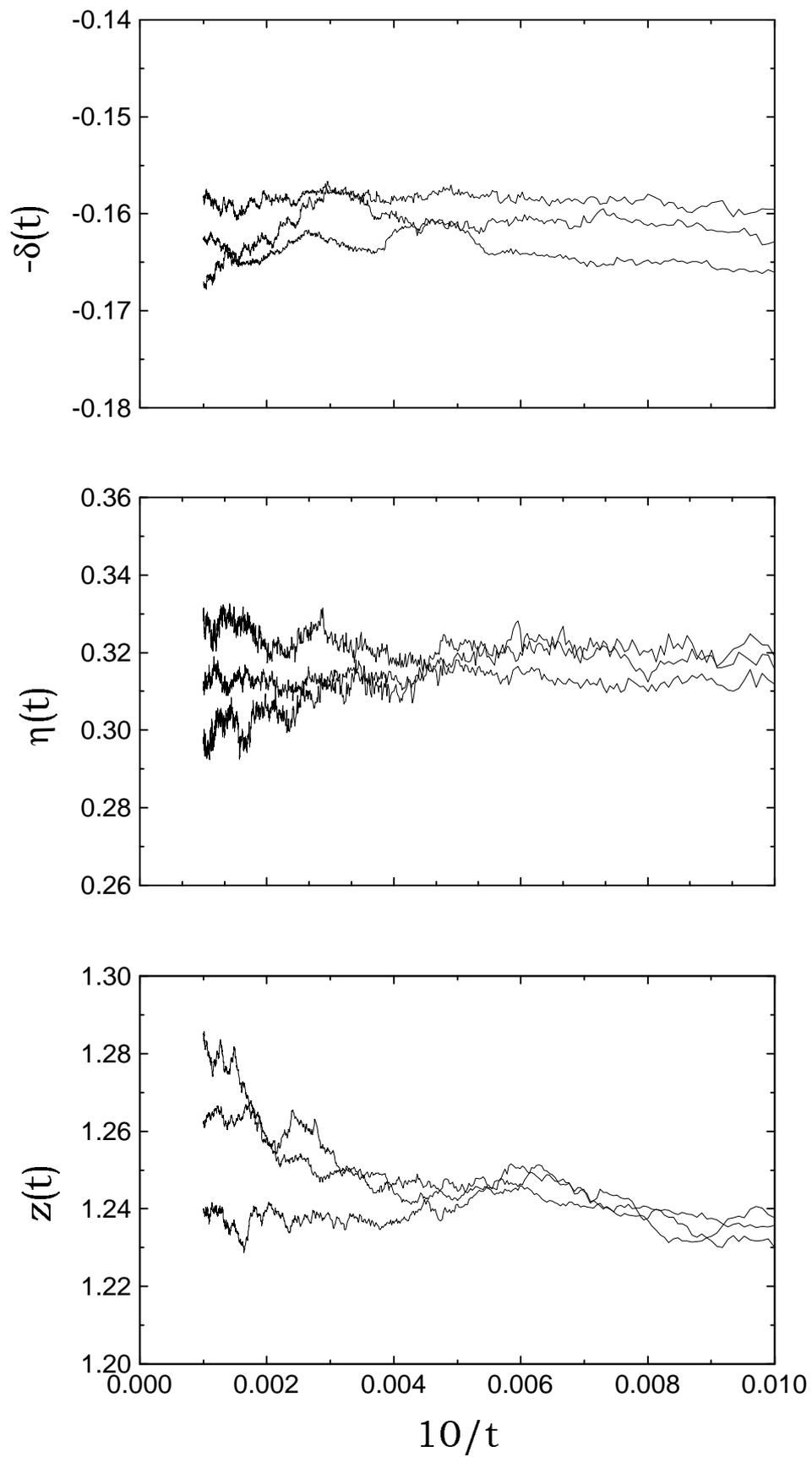
\begin{figure}
\caption{Plots of the effective exponents against $10/t$ for the 
defect dynamics of the asymmetric IMD model with $h=0.5$.
Three curves from top to bottom in each panel correspond to
$p=0.4132$, $0.4138$, and $0.4144$.}
\end{figure}

\begin{figure}
\caption{Evolutions of the asymmetric IMD interface dynamics for 
$(a)$ $p<p_c$, $(b)$ $p=p_c$, and $(c)$ $p>p_c$. 
The region of the preferred (unpreferred) absorbing state 
is shown in black (grey). The active sites are
represented by white pixels.}
\end{figure}

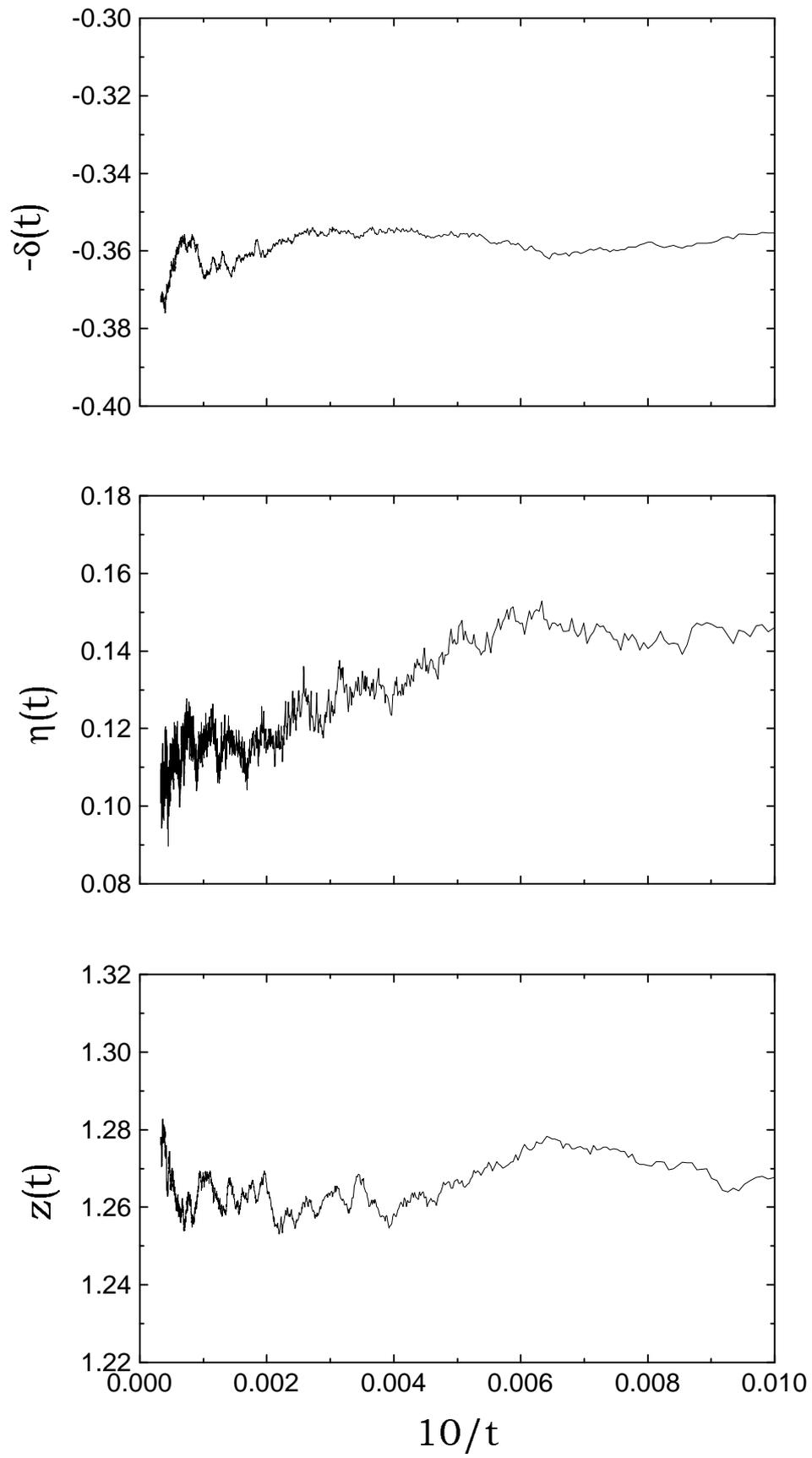
\begin{figure}
\caption{Plots of the effective exponents against $10/t$ for the 
type-C interface dynamics of the asymmetric IMD model for $h=0.5$
at criticality.}
\end{figure}

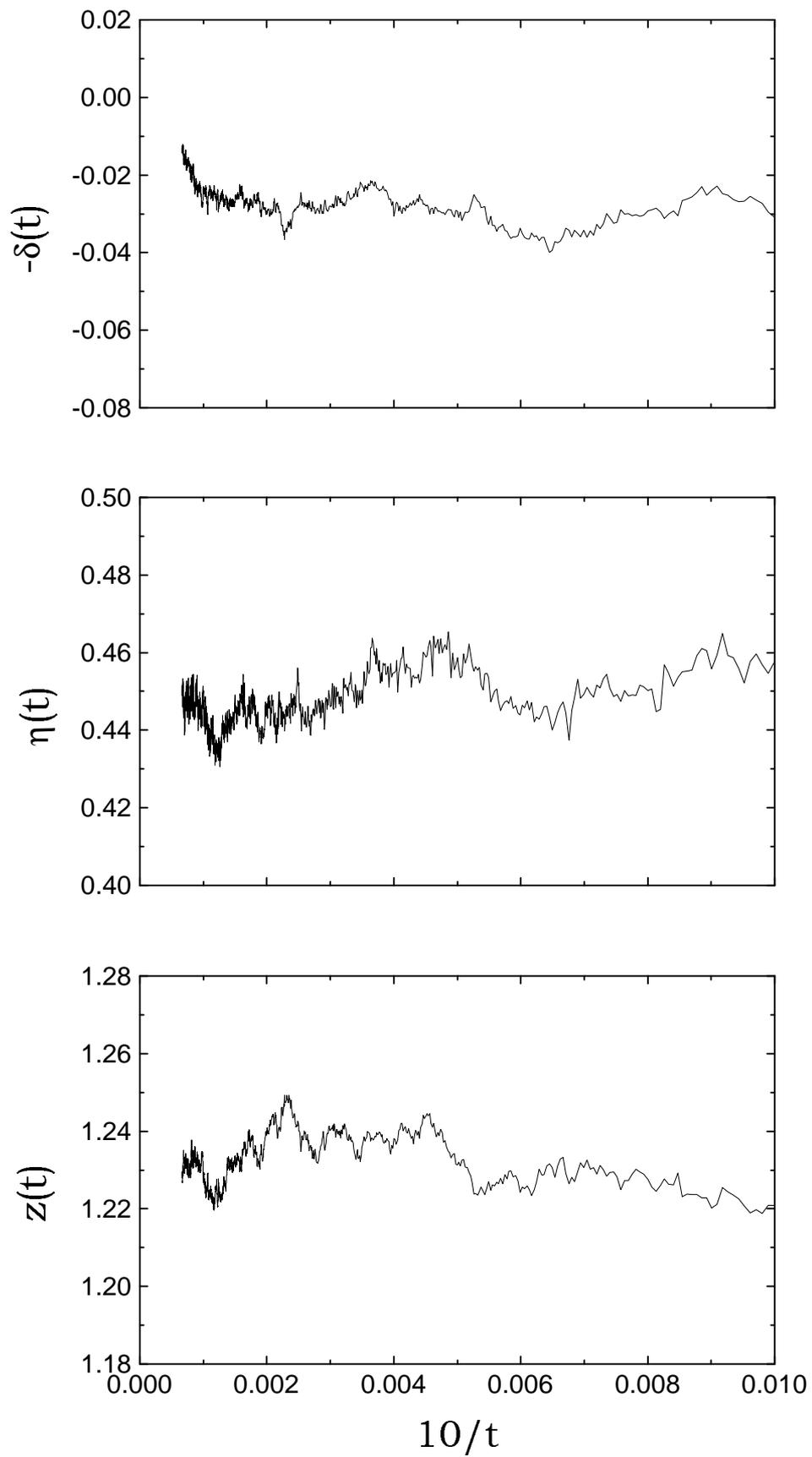
\begin{figure}
\caption{Plots of the effective exponents against $10/t$ for the 
type-D interface dynamics of the asymmetric IMD model for $h=0.5$
at criticality.}
\end{figure}

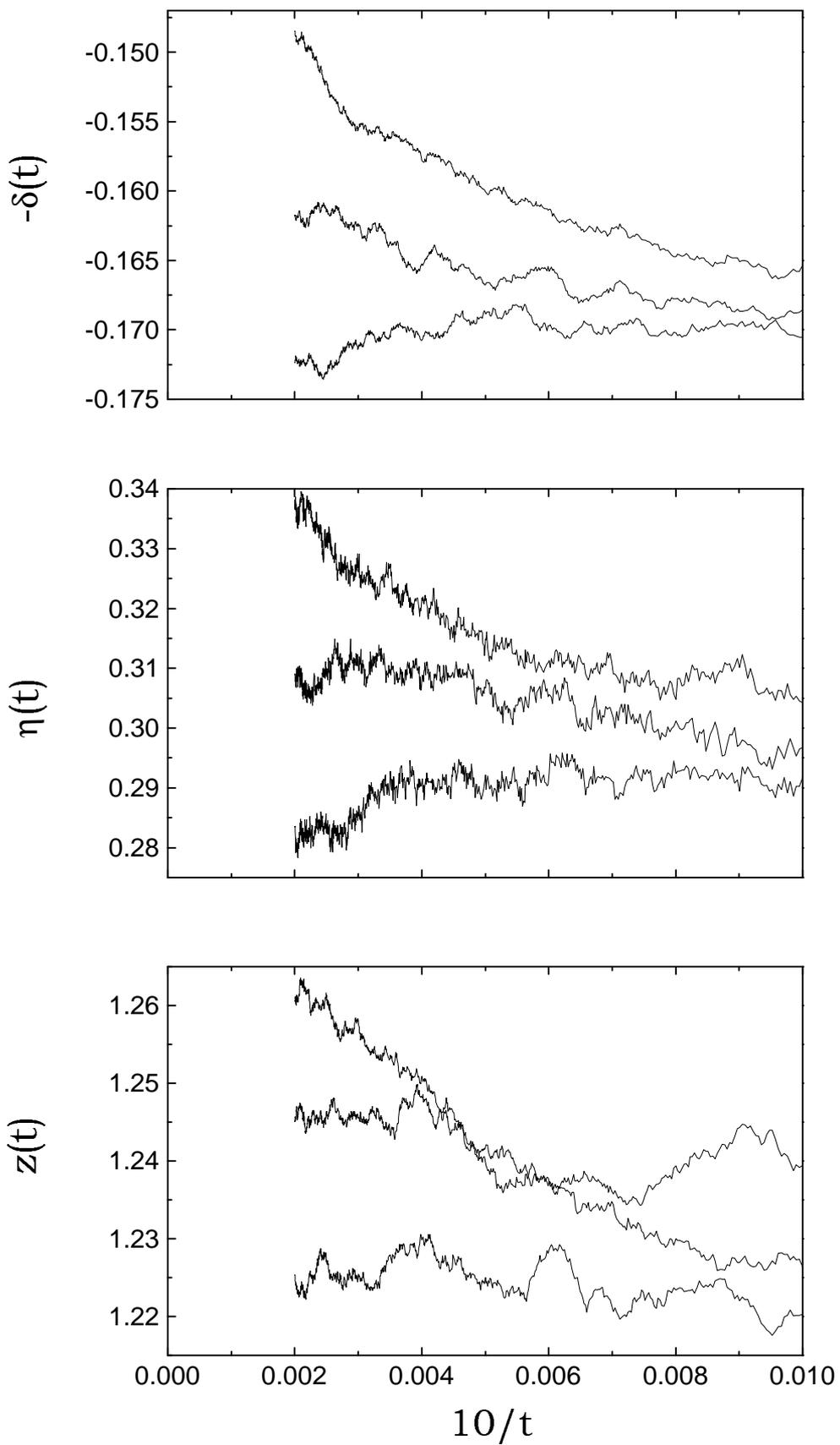
\begin{figure}
\caption{Plots of the effective exponents against $10/t$ for the 
defect dynamics of the asymmetric PCA A model with $h=0.1$.
Three curves from top to bottom in each panel correspond to
$p=0.2443$, $0.2435$, and $0.2427$.}
\end{figure}

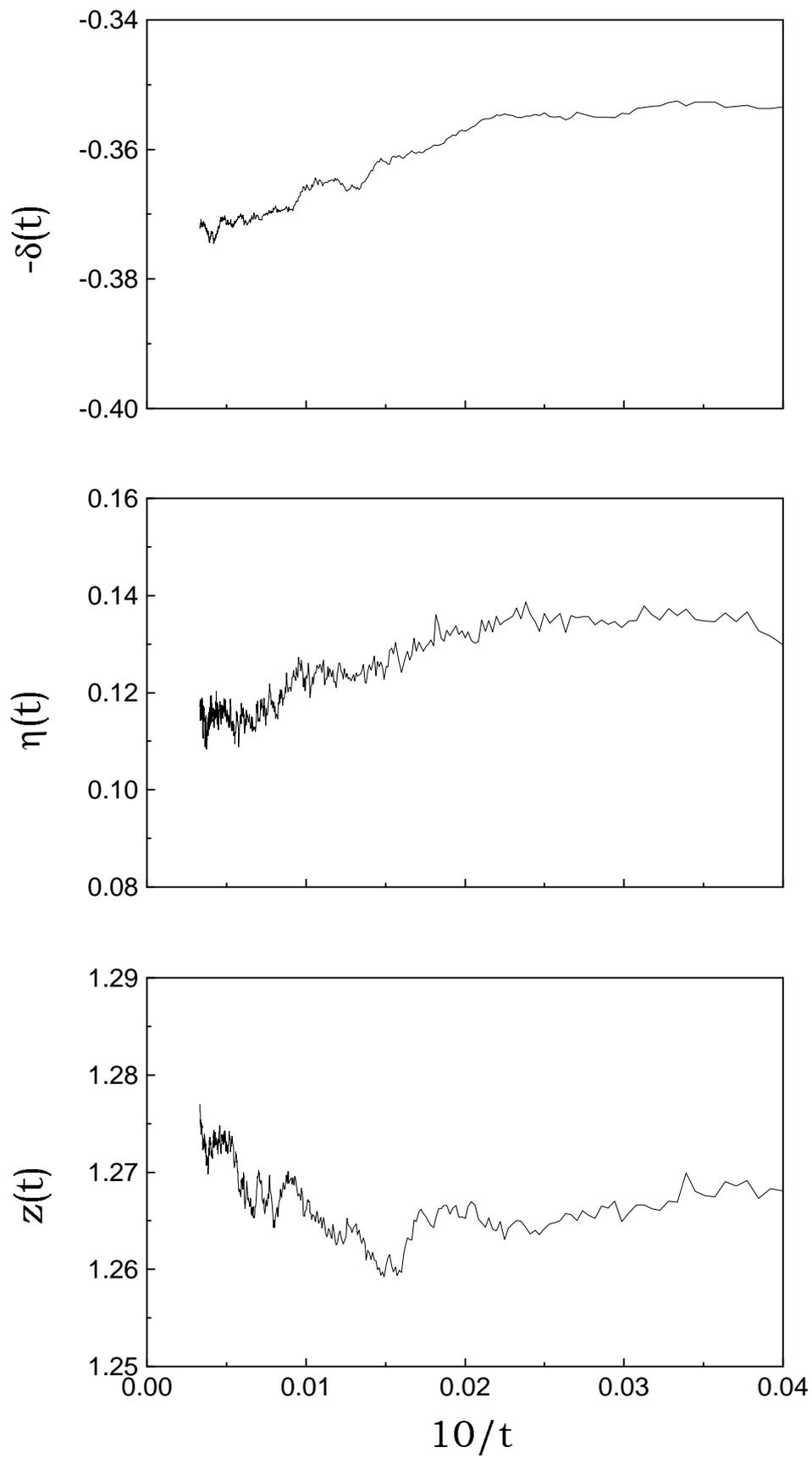
\begin{figure}
\caption{Plots of the effective exponents against $10/t$ for the 
type-C interface dynamics of the asymmetric PCA A model with $h=0.1$
at criticality.}
\end{figure}

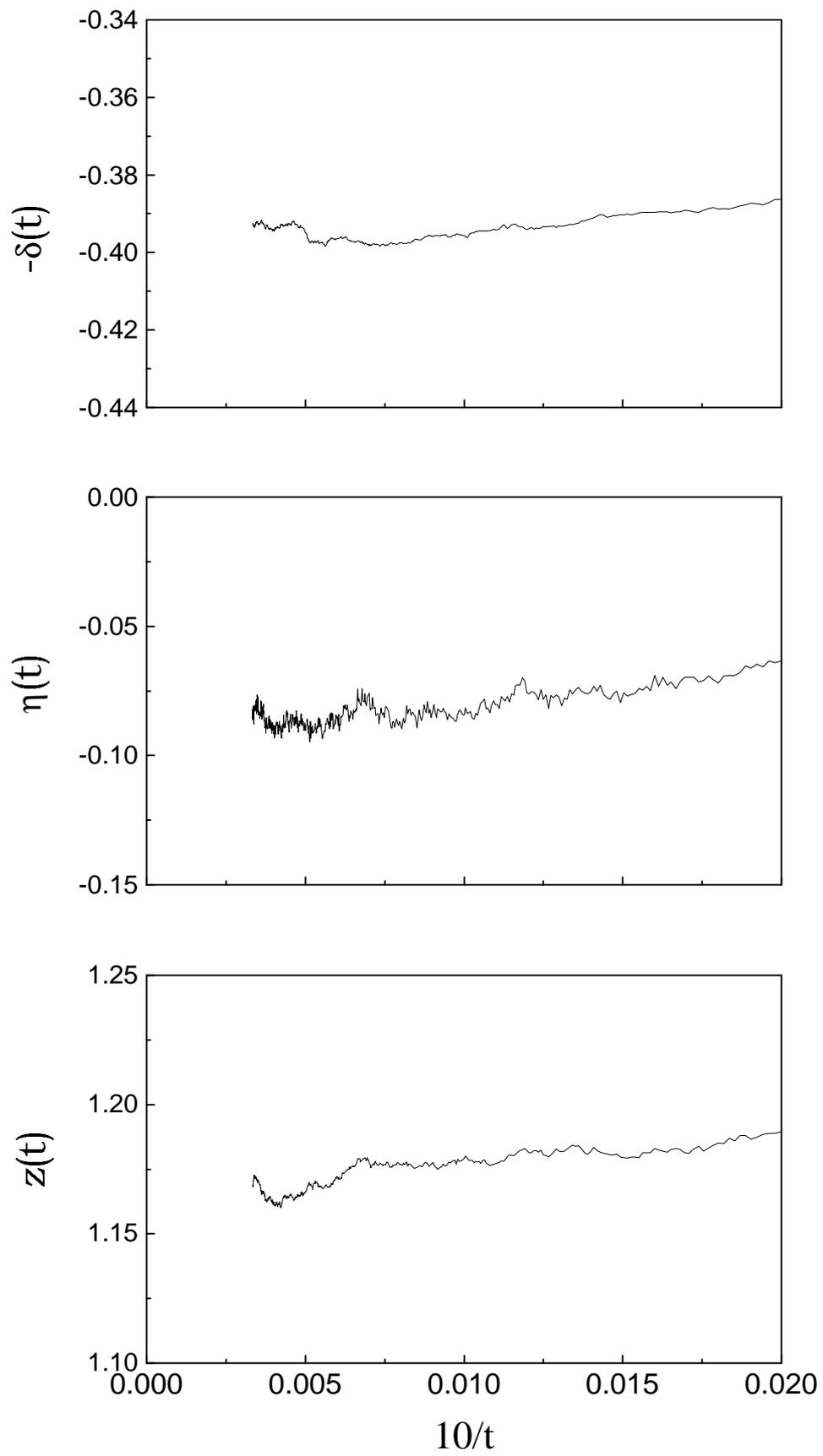
\begin{figure}
\caption{Plots of the effective exponents against $10/t$ for the 
type-C interface dynamics of the symmetric PCA A model
at criticality.}
\end{figure}

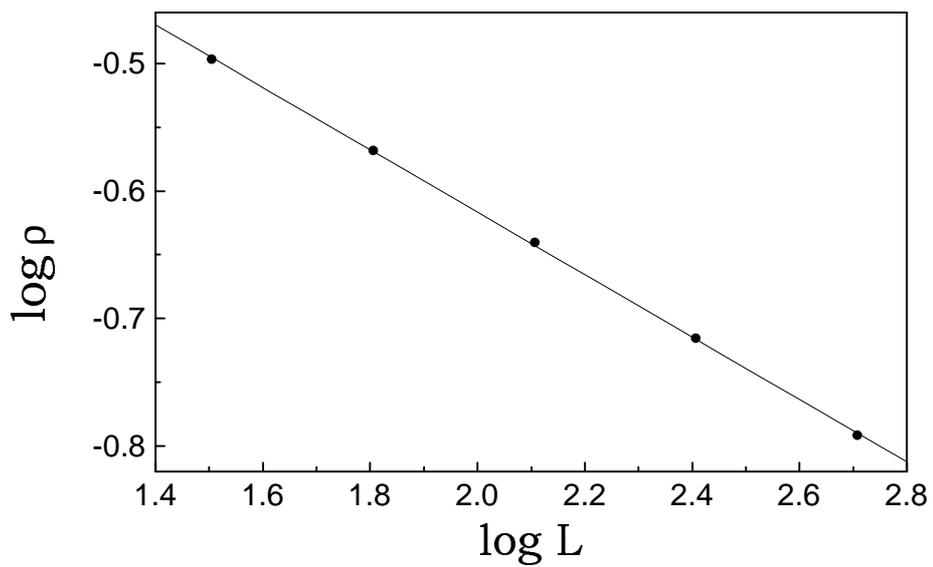
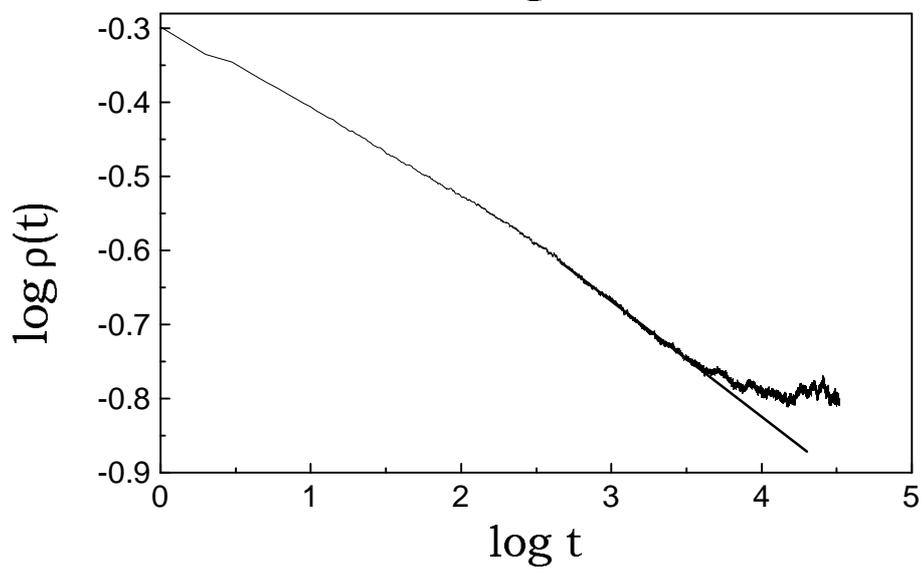
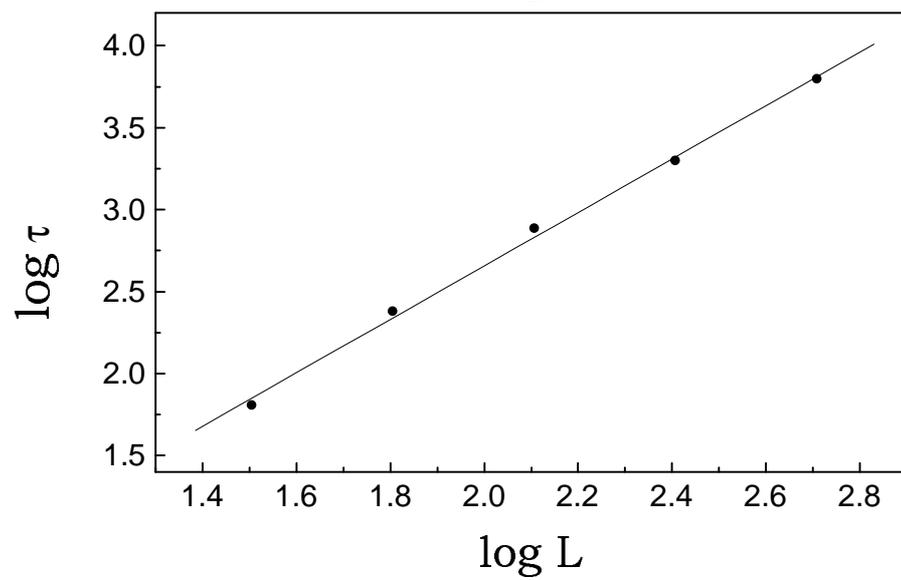
\begin{figure}
\caption{The kink density in the quasisteady state
against the system size, time dependence of the kink density
at $L=512$, and size dependence of the characteristic time are plotted 
for the asymmetric PCA A model with $h=0.1$
at criticality.
The solid lines are of slope $-0.245$, $-0.155$,
and $1.63$ from top to bottom.}
\end{figure}

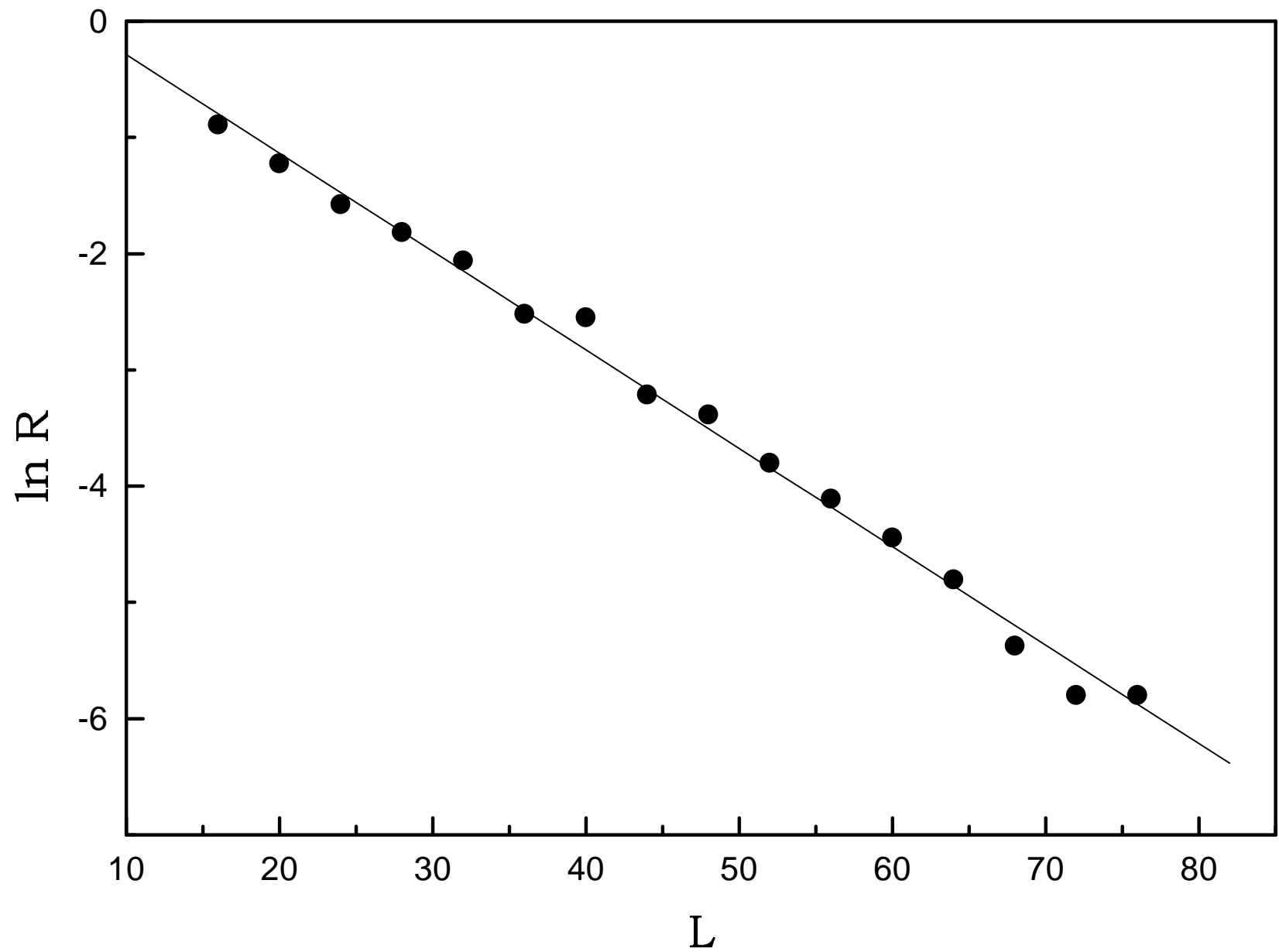
\begin{figure}
\caption{The semi-logarithmic plot for the raio $R$ against system size
$L$. The solid is $R=1.44 \exp{(-L/12.9)}$.}
\end{figure}

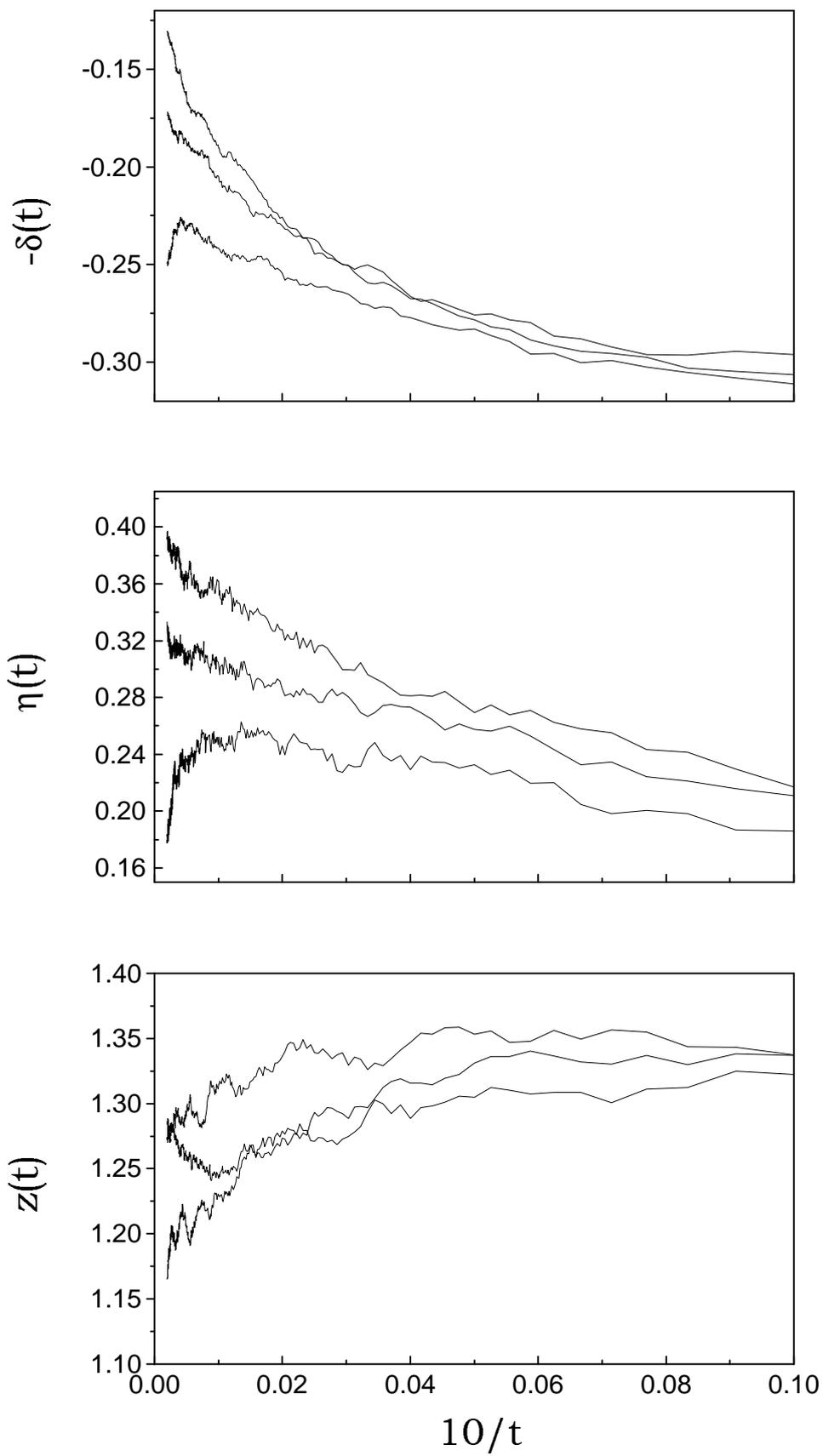
\begin{figure}
\caption{Plots of the effective exponents against $10/t$ for the 
defect dynamics of the asymmetric NKI model with $h=0.1$.
Three curves from top to bottom in each panel correspond to
$p=0.18$, $0.19$, and $0.20$.}
\end{figure}

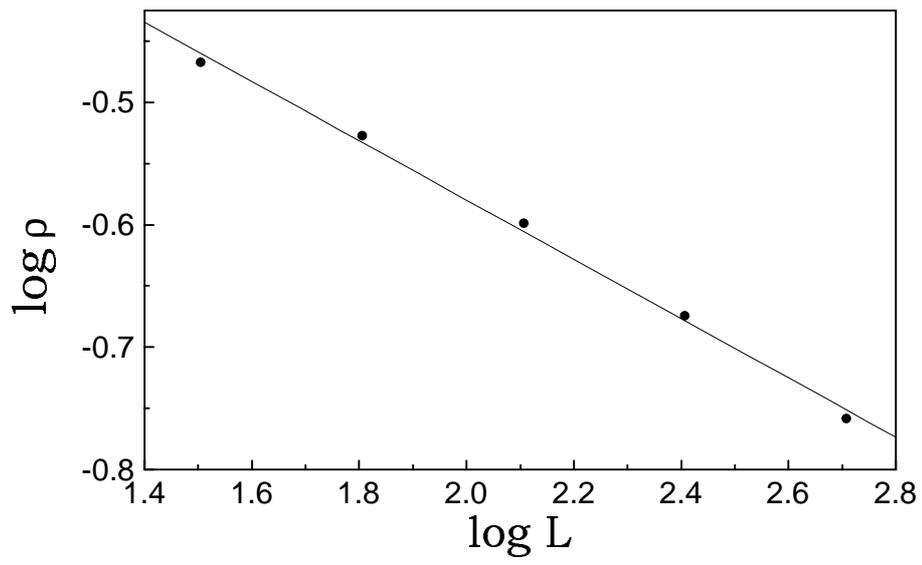
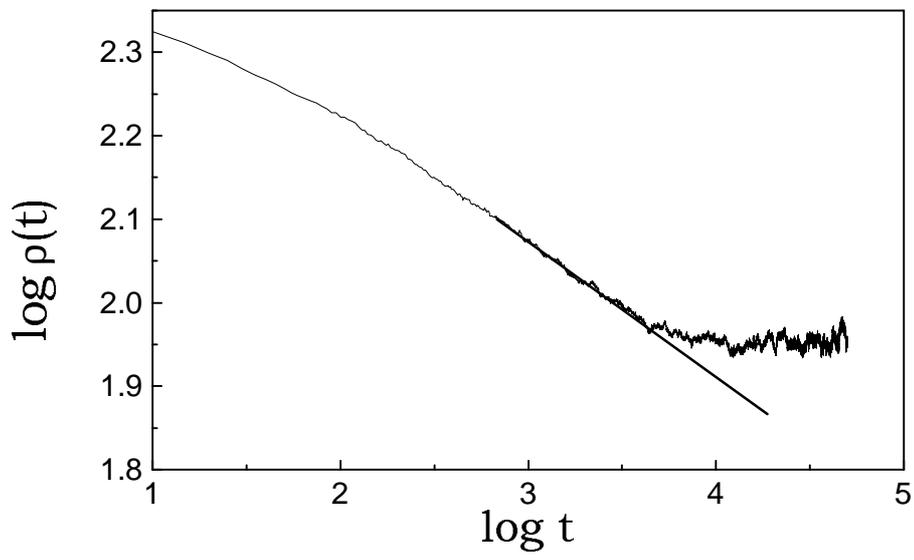
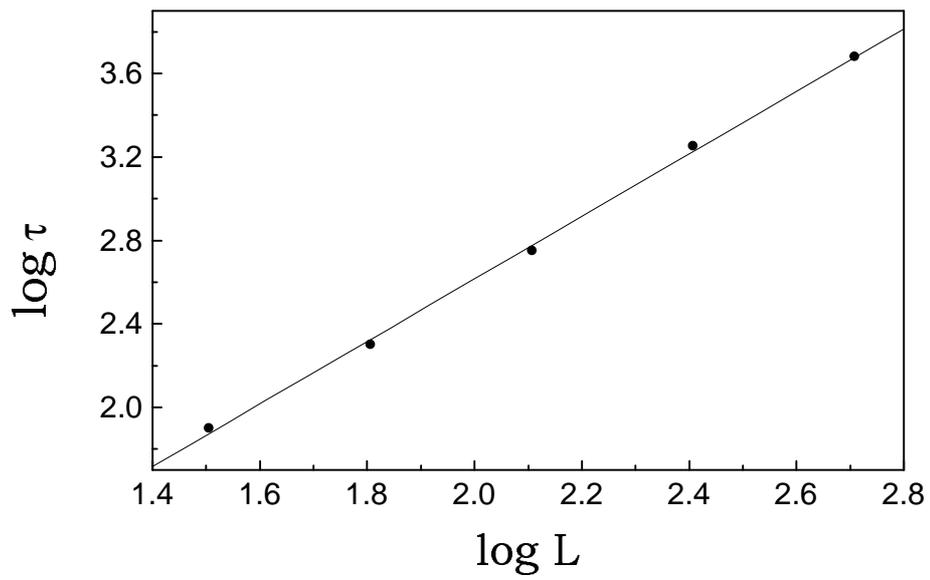
\begin{figure}
\caption{The kink density in the quasisteady state
against the system size, time dependence of the kink density
at $L=512$, and size dependence of the characteristic time are plotted 
for the asymmetric NKI model with $h=0.1$
at criticality. The solid lines are of slope $-0.24$, $-0.155$,
and $1.50$ from top to bottom.}
\end{figure}

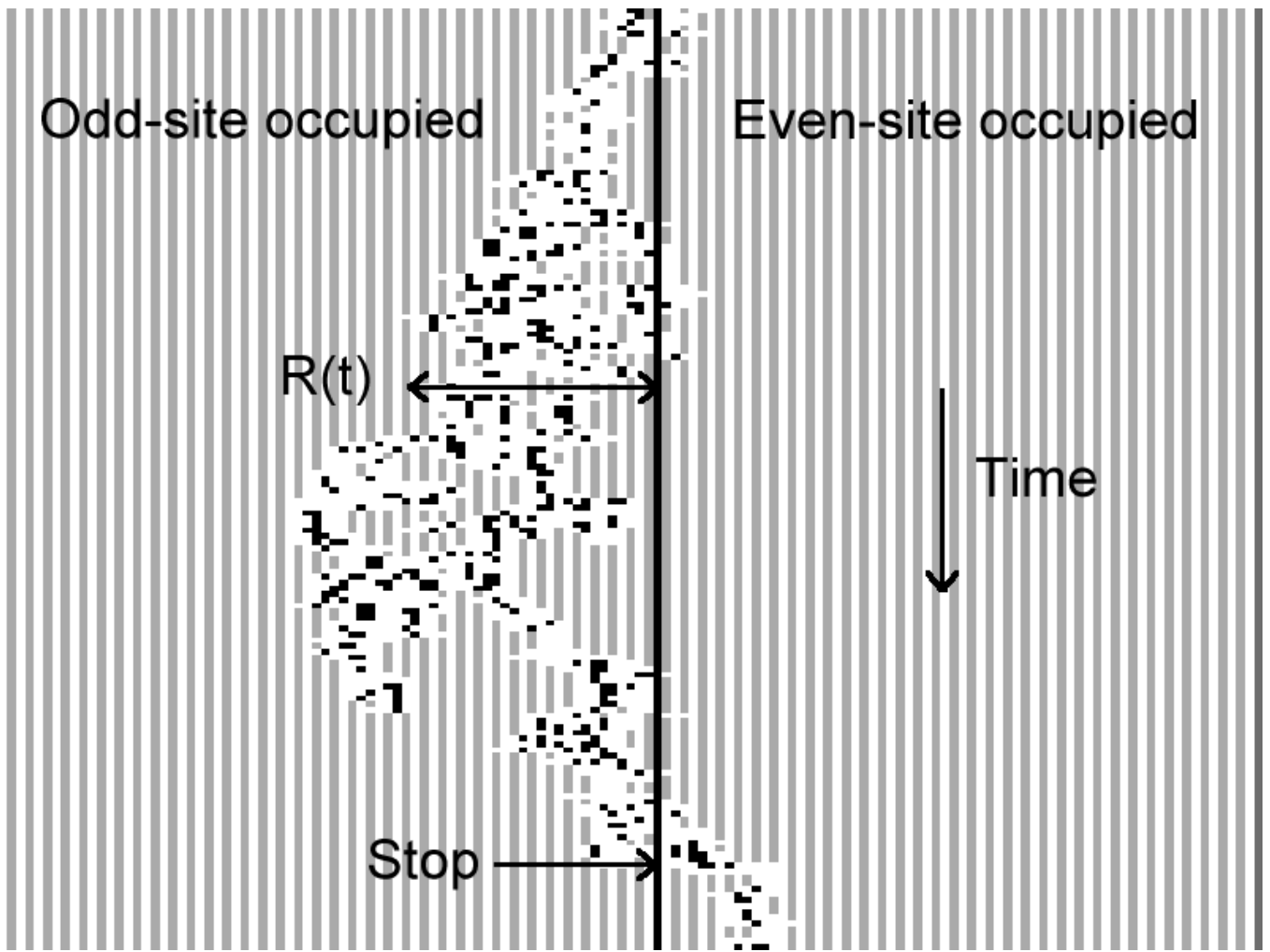
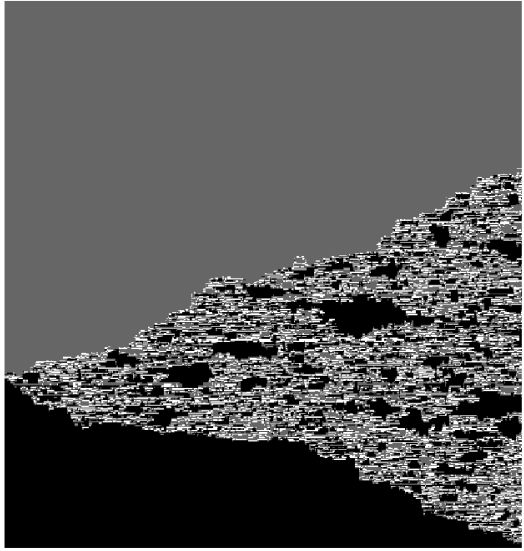
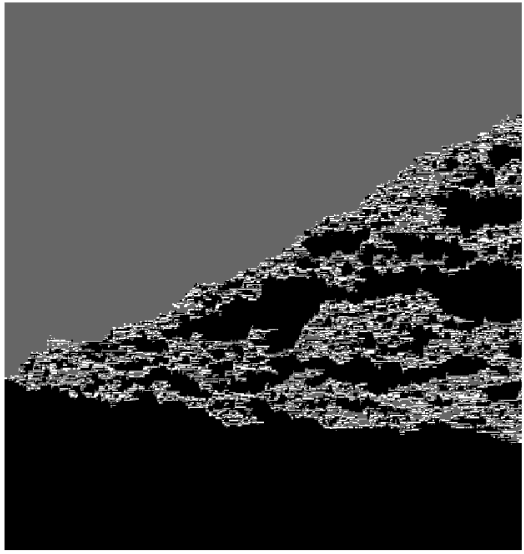
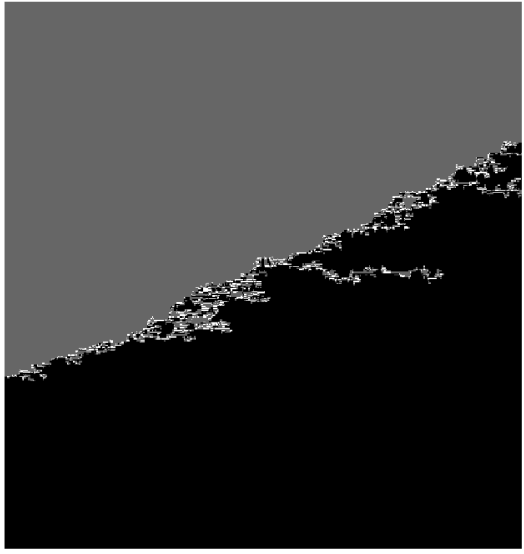
\end{document}